\RequirePackage{fix-cm}
\documentclass[twocolumn, epjc3, comma, sort&compress, natbib]{svjour3}
\bibpunct{[}{]}{,}{n}{}{,} 

\journalname{Eur. Phys. J. C}

\usepackage{latexsym}
\usepackage{amsmath}
\usepackage{amssymb}
\usepackage{amsfonts}
\usepackage{CJKutf8}

\usepackage[mathscr,scaled=1.15]{urwchancal}
\DeclareFontFamily{OT1}{pzc}{}
\DeclareFontShape{OT1}{pzc}{m}{it}%
{<-> s * [1.15] pzcmi7t}{}
\DeclareMathAlphabet{\mathpzc}{OT1}{pzc}{m}{it}

\usepackage{color}

\usepackage{supertabular}
\usepackage{placeins}
\usepackage{epsfig}
\usepackage{graphicx}

\definecolor{purple}{rgb}{0.5,0,0.5}
\definecolor{blue}{rgb}{0.0,0,0.9}
\definecolor{prdblue}{rgb}{0.133,0.118,0.498}
\usepackage[colorlinks=true, pdfstartview=FitV, linkcolor=prdblue, citecolor= prdblue, urlcolor=prdblue]{hyperref}

\begin{document}
\begin{CJK*}{UTF8}{gbsn}

\title{$\,$\\[-6ex]\hspace*{\fill}{\normalsize{\sf\emph{Preprint nos}.\
NJU-INP 096/25}}\\[1ex]
Distribution Functions of a Radially Excited Pion}

\author{Z.-N. Xu (徐珍妮)\thanksref{UHe}%
    $^{\href{https://orcid.org/0000-0002-9104-9680}{\textcolor[rgb]{0.00,1.00,0.00}{\sf ID}},}$
    %
\and
    Z.-Q. Yao (姚照千)\thanksref{UHe,UPO}
       $\,^{\href{https://orcid.org/0000-0002-9621-6994}{\textcolor[rgb]{0.00,1.00,0.00}{\sf ID}}}$
\and
    D. Binosi\thanksref{ECT}%
    $\,^{\href{https://orcid.org/0000-0003-1742-4689}{\textcolor[rgb]{0.00,1.00,0.00}{\sf ID}}}$
\and
    \\M.~Ding (丁明慧)\thanksref{NJU,INP}%
    $\,^{\href{https://orcid.org/0000-0002-3690-1690}{\textcolor[rgb]{0.00,1.00,0.00}{\sf ID}}}$
\and
    C. D. Roberts\thanksref{NJU,INP}%
       $^{\href{https://orcid.org/0000-0002-2937-1361}{\textcolor[rgb]{0.00,1.00,0.00}{\sf ID}},}$
\and
    J. Rodr\'iguez-Quintero\thanksref{UHe}%
       $^{\href{https://orcid.org/0000-0002-1651-5717}{\textcolor[rgb]{0.00,1.00,0.00}{\sf ID}},}$
}

\authorrunning{Zhen-Ni Xu \emph{et al}.} 

\institute{Dpto.~Ciencias Integradas, Centro de Estudios Avanzados en Fis., Mat. y Comp., \\
\hspace*{0.5em}Fac.~Ciencias Experimentales, \href{https://ror.org/03a1kt624}{Universidad de Huelva}, E-21071 Huelva, Spain
\label{UHe}
\and
Dpto. Sistemas F\'isicos, Qu\'imicos y Naturales, Univ.\ \href{https://ror.org/02z749649}{Pablo de Olavide}, E-41013 Sevilla, Spain
\label{UPO}
\and
European Centre for Theoretical Studies in Nuclear Physics
            and Related Areas  (\href{https://ror.org/01gzye136}{ECT*}), \\
            \hspace*{0.5em}Villa Tambosi, Strada delle Tabarelle 286, I-38123 Villazzano (TN), Italy
\label{ECT}
\and
School of Physics, \href{https://ror.org/01rxvg760}{Nanjing University}, Nanjing, Jiangsu 210093, China \label{NJU}
\and
Institute for Nonperturbative Physics, \href{https://ror.org/01rxvg760}{Nanjing University}, Nanjing, Jiangsu 210093, China \label{INP}
\\[1ex]
Email:
\href{mailto:zhaoqian.yao@dci.uhu.es}{zhaoqian.yao@dci.uhu.es} (ZQY);
\href{mailto:mhding@nju.edu.cn}{mhding@nju.edu.cn} (MD);
\href{mailto:cdroberts@nju.edu.cn}{cdroberts@nju.edu.cn} (CDR)
            }

\date{2025 January 22}

\maketitle

\end{CJK*}

\begin{abstract}
A nonperturbatively-improved, symmetry-\linebreak preserving approximation to the quantum field equations relevant in calculations of meson masses and interactions is used to deliver predictions for all distribution functions (DFs) of the ground state pion, $\pi_0$, and its first radial excitation, $\pi_1$, \emph{viz}.\ valence, glue, and sea.
Regarding Mellin moments of the valence DFs, the $m=0,1$ moments in both states are identical; but for each $m\geq 2$, that in the $\pi_0$ is greater than its partner in the $\pi_1$.
Working with such information, pointwise reconstructions of the hadron-scale $\pi_{0,1}$ valence DFs are developed.
The predicted $\pi_0$ valence DF is consistent with extant results.
The $\pi_1$ valence DF is novel: it possesses three-peaks, with the central maximum partnered by secondary peaks on either side, each separated from the centre by a zero: the zeroes lie at $x\approx 0.2,0.8$ and the secondary peaks at $x\approx 0.1,0.9$.
Evolution to $\zeta =3.2\,$GeV, a typical scale for nonperturbative calculations, is accomplished using an evolution scheme for parton DFs that is all-orders exact.
At this higher scale, differences between the $\pi_{0,1}$ valence DFs remain significant, but analogous differences between glue and sea DFs are far smaller.
This analysis shows that, owing to constraints imposed by chiral symmetry and the pattern by which it is broken in Nature, there are noticeable differences between the structural properties of the pion ground state and its radial excitations.
\end{abstract}

\section{Introduction}
The pion is a bound state seeded by a light quark and light antiquark, \emph{e.g}., $\pi^+ = u \bar d$.  It is also Nature's most fundamental Nambu-Goldstone boson.  This dichotomy is reconciled by dynamical chiral symmetry breaking (DCSB) \cite{Maris:1997hd, Brodsky:2012ku, Qin:2014vya, Horn:2016rip}, itself a corollary of emergent hadron mass (EHM) \cite{Roberts:2021nhw, Binosi:2022djx, Ding:2022ows, Roberts:2022rxm, Ferreira:2023fva, Carman:2023zke, Raya:2024ejx}.
The reconciliation is readily seen by exploiting the axialvector Ward-Green-Takahashi\linebreak identity and associated Bethe-Salpeter equations in\linebreak quantum chromodynamics (QCD).  For instance, one finds therewith \cite{Maris:1997hd, Brodsky:2012ku, Qin:2014vya}:
\begin{equation}
f_\pi m_\pi^2 = (m_u^\zeta + m_d^\zeta) r_\pi^\zeta\,,
\label{GMOR}
\end{equation}
where $m_\pi$ is the pion mass; $f_\pi$ is the pseudovector projection of the pion's Poincar\'e-covariant  wave function onto the origin in configuration space, \emph{i.e}., the pion leptonic decay constant; $r_\pi^\zeta$ is its pseudoscalar analogue; and $m_{u,d}^\zeta$ are the quark current masses introduced by Higgs boson couplings into QCD.  Each term on the right-hand side of Eq.\,\eqref{GMOR} depends on the resolving scale, $\zeta$, but the product is scale invariant.

With Eq.\,\eqref{GMOR}, one recovers what is commonly called the Gell-Mann--Oakes--Renner relation \cite{GellMann:1968rz}.
As an order parameter for DCSB, $f_\pi$ remains nonzero as Higgs boson couplings are removed.
In fact, $f_\pi^0 \approx 0.088\,$GeV \cite{Gasser:1983yg}.
(The superscript ``$0$'' denotes chiral limit.)
Consequently, Eq.\,\eqref{GMOR} entails that the mass of the ground-state pion vanishes in the chiral limit.

Since $f_\pi^0 \neq 0$, then the chiral-limit pion Bethe-Sal\-peter amplitude (hence, wave function) is also nonzero.  To be explicit, recall that the Poincar\'e-covariant amplitude for a light pseudoscalar meson has the following form (isospin symmetry is assumed, and flavour and colour matrices are suppressed) \cite{LlewellynSmith:1969az}:
\begin{align}
\Gamma_{\pi}(k;&P)  = \gamma_5
\left[ i E_{\pi}(k;P)
+ \gamma\cdot P F_{\pi}(k;P) \right. \nonumber \\
& \left. + \gamma\cdot k k\cdot P G_{\pi}(k;P)
+ \sigma_{\mu\nu} k_\mu P_\nu H_{\pi}(k;P)
\right]\,, \label{BSAmp}
\end{align}
where $P$ is the total momentum of the bound state and $k$ is the relative momentum between the valence quark + antiquark pair.
In the chiral limit, the leading term in the pion Bethe-Salpeter amplitude, $E_\pi$, becomes indistinguishable from the scalar piece of the dressed quark self-energy \cite{Maris:1997hd}:
\begin{equation}
f_\pi^0 E_\pi^0(k,0;\zeta) = B^0(k;\zeta)\,.
\label{gtrE}
\end{equation}

The crucial link between the pseudoscalar two-body and quark one-body problems, expressed by Eq.\,\eqref{gtrE}, is the most fundamental expression of Goldstone's theorem in QCD.  No model that fails to preserve it can pretend to a veracious description of pion properties, whether or not it achieves a massless pion in some limit.

It has long been realised that Eq.\,\eqref{GMOR} is a special case of a more general formula, which is valid for every pseudoscalar meson \cite{Dominguez:1976ut, Holl:2004fr, Holl:2005vu, Bhagwat:2006xi, Ballon-Bayona:2014oma, Jiang:2015paa}:
\begin{equation}
f_{\pi_n} m_{\pi_n}^2 = (m_u^\zeta + m_d^\zeta) r_{\pi_n}^\zeta\,,
\label{GMORn}
\end{equation}
where $\pi_n$ is radial excitation $n$ of the pion ground-state, labelled $\pi_0$ hereafter.
(We continue to discuss pion-like systems, but the result is not limited to such bound states.)
Calculations and experiment show that pion radial excitations exist and are very massive; hence, in the chiral limit:
\begin{equation}
\forall n \geq 1 \; | \; f_{\pi_n}^0 \equiv 0 \,.
\label{fpin0}
\end{equation}
Consistent with this prediction, an array of calculations at physical quark current masses indicate $|f_{\pi_1}| < 10\,$MeV \cite{Holl:2004fr, Holl:2005vu, Bhagwat:2006xi, Diehl:2001xe, McNeile:2006qy, Ballon-Bayona:2014oma, Xu:2022kng}.  Our prediction is
\begin{equation}
f_{\pi_1} = 7.8(5)\,{\rm MeV}.
\label{fpin1}
\end{equation}

Hereafter, we focus on the ground state pion and its first radial excitation.
Plainly, given Eq.\,\eqref{GMORn}, then Eq.\,\eqref{gtrE} is not valid for the $\pi_1$.
In fact, the Poincar\'e-covariant Bethe-Salpeter amplitudes (hence, wave functions) for $\pi_0$ and $\pi_1$ are very different.
This is readily highlighted by considering the light-front projections of their respective Bethe-Salpeter wave functions, as done elsewhere \cite{Li:2016dzv}.  The functions thus obtained are the meson distribution amplitudes (DA).

\begin{figure}[t]
\centerline{%
\includegraphics[clip, width=0.43\textwidth]{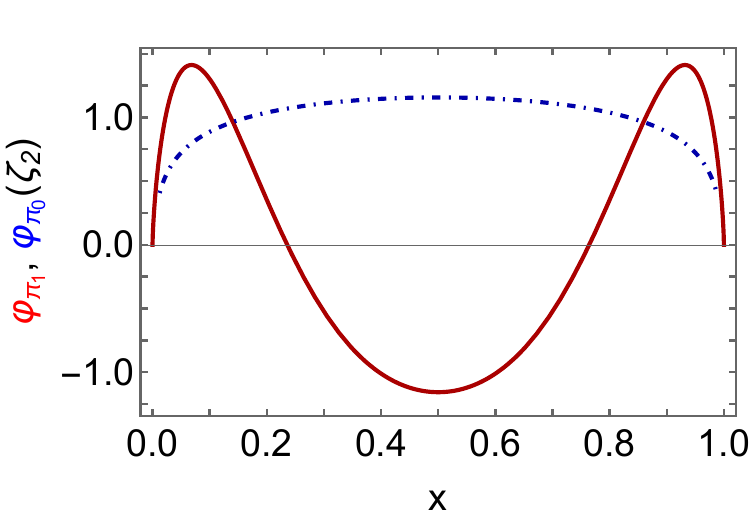}}
\caption{\label{FigDA1}
Leading-twist two-quasiparticle distribution amplitude of the pion ground state (dot-dashed blue) and its first radial excitation (red solid), calculated at resolving scale $\zeta = \zeta_2 = 2\,$GeV \cite{Li:2016dzv}.
}
\end{figure}

The comparison from Ref.\,\cite{Li:2016dzv} is redrawn in Fig.\,\ref{FigDA1}.
In this figure, $\varphi_{\pi_0}(x;\zeta_2)$ is dimensionless; and in order to make the comparison simple, we have rescaled $\varphi_{\pi_1}(x;\zeta_2)$ so, in magnitude, its $x=1/2$ value matches that of $\varphi_{\pi_0}(x;\zeta_2)$.
The ground state DA is a broad concave function, whose dilation -- confirmed by numerous other analyses, \emph{e.g}., Refs.\,\cite{Roberts:2021nhw, LatticeParton:2022zqc, Li:2022qul, Gao:2022vyh, Xing:2023wuk, Lu:2023yna, Alexandrou:2024zvn} -- is a consequence and signal of EHM \cite{Chang:2013pq}.
Owing to EHM, $\varphi_{\pi_1}(x;\zeta_2)$ is also dilated; but the most striking features are the two zeroes, which emphasise the character of $\pi_1$ as a radial excitation, and the fact that, expressing Eqs.\,\eqref{fpin0}, \eqref{fpin1}, the domains of positive and negative support are almost perfectly matched. (There are $2n$ zeroes in $\varphi_{\pi_n}$.)

No model of pseudoscalar meson structure can claim a connection with QCD unless it veraciously reproduces the symmetry driven outcomes described above, \emph{viz}.\ achieves them and does so without tuning one or another model aspect.  Perplexingly, these tight symmetry constraints are often overlooked.

Herein, we aim to determine the impact of the striking differences between the wave functions of ground and radially excited pseudoscalar mesons on their parton distribution functions (DFs).
In Sect.\,\ref{secDFformula}, we introduce the formula used to calculate hadron-scale, $\zeta_{\cal H}$, valence quark DFs in a pion-like system, highlight some features of these DFs, and introduce the all-orders evolution scheme used to obtain all DFs at scales above $\zeta_{\cal H}$.
Section~\ref{ELequations} describes the quantum field equations used to calculate the Schwinger functions that appear in the DF formula and explains the kernels used therein.  The efficacy of our formulation of the meson bound-state problem is illustrated via predictions for an array of meson masses and leptonic decay constants.
The methods used to calculate Mellin moments of the hadron-scale valence quark DFs are explained in Sect.\,\ref{SecPred}, which also includes the results obtained therewith.
Section~\ref{DFRecon} describes our procedure for reconstructing the pointwise behaviour of a DF from a collection of its Mellin moments and subsequently reports comparisons between the hadron-scale $\pi_{0,1}$ valence DFs.
In Sect.\,\ref{SecAOE}, evolution to scales above $\zeta_{\cal H}$ is explained and used to deliver predictions for and comparisons between all DFs in the ground-state pion and its first radial excitation, \emph{viz}.\ valence, glue, and separated four-flavour sea.
A summary and perspective are provided in Sect.\,\ref{epilogue}.

\section{Pseudoscalar Meson DFs and Evolution}
\label{secDFformula}
The parton distribution function associated with a valence degree of freedom in a pseudoscalar meson can be calculated using the following formula \cite{Chang:2014lva, Ding:2019qlr, Ding:2019lwe}:
\begin{align}
{\mathpzc q}^\pi(x;\zeta_{\cal H})  &= N_c {\rm tr}\! \int \frac{d^4k}{(2\pi)^4}\! \delta_{\mathpzc n}^{x}(k_\eta) \Gamma_\pi^P(k_{\bar\eta\eta};\zeta_{\cal H})\, S(k_{\bar\eta};\zeta_{\cal H})
\nonumber \\
& \times \{n\cdot\frac{\partial}{\partial {k_\eta}} \left[ \Gamma_\pi^{-P}(k_{\eta\bar\eta};\zeta_{\cal H}) S(k_\eta;\zeta_{\cal H}) \right]\}\,,
\label{qFULL}
\end{align}
wherein
$x$ is the light-front fraction of the bound-state total momentum, $P$, carried by the struck degree of freedom;
projection onto the light front is enforced by
$\delta^x_{\mathpzc n}=\delta({\mathpzc n}\cdot k_\eta - x {\mathpzc n}\cdot P)$, with ${\mathpzc n}$ a light-like four-vector, ${\mathpzc n}^2=0$ and ${\mathpzc n}\cdot P = -m_\pi$ in the meson rest frame;
$S$ is the dressed quark propagator;
and $k_\eta=k+\eta P$, $k_{\bar\eta}=k-(1-\eta)P$, $k_{\eta\bar\eta}=(k_\eta+k_{\bar\eta})/2$, $\eta\in [0,1]$, is the relative momentum between the valence degrees of freedom.
Owing to Poincar\'e invariance, no observable can legitimately depend on $\eta$; so, for pion-like systems, it is common to choose $\eta=1/2$, which is implicit in Eq.\,\eqref{BSAmp}.
Moreover, it follows from the light-front wave function overlap representation of unpolarised generalised parton DFs and their forward-limit connection to collinear DFs \cite{Diehl:2003ny} that, $\forall n \geq 0$, ${\mathpzc q}^{\pi_n}(x;\zeta_{\cal H}) $ is a non-negative function on $x\in [0,1]$.

Some significant features of Eq.\,\eqref{qFULL} are worth highlighting.
(\emph{i})  Irrespective of the approximation used to solve all quantum field equations relevant to the meson bound state problem, so long as it is symmetry preserving, then canonical normalisation of the Bethe-Salpeter amplitude \cite{Nakanishi:1969ph} ensures baryon number conservation:
\begin{equation}
\int_0^1 dx\, {\mathpzc q}^\pi(x;\zeta_{\cal H}) = 1\,.
\end{equation}
%
(\emph{ii}) Owing to the symmetric character of the integrand, which ensures that neither valence degree of freedom is favoured:
\begin{subequations}
\label{valencemom}
\begin{align}
                & {\mathpzc q}^\pi(x;\zeta_{\cal H}) = {\mathpzc q}^\pi(1-x;\zeta_{\cal H})
                \label{valencemoma}\\
\Rightarrow & \int_0^1 dx\, x {\mathpzc q}^\pi(x;\zeta_{\cal H}) = \tfrac{1}{2}.
\end{align}
\end{subequations}
This is the statement that valence degrees of freedom carry all the momentum of the bound state at the had\-ron scale, $\zeta_{\cal H}$.

The empirical inference of parton DFs is only possible at scales $\zeta > m_p$ ($m_p$ is the proton mass), whereat factorisation theorems are valid \cite[Chap.\,4.3]{Ellis:1991qj}.  At such scales, valence degrees of freedom do not carry all the hadron's light-front momentum.  Thus DFs obtained\linebreak from Eq.\,\eqref{qFULL} must be scale-evolved in order for comparisons to be made.  We use the all-orders (AO) scheme, whose development began with Refs.\,\cite{Ding:2019qlr, Ding:2019lwe} and which is succinctly explained in Ref.\,\cite{Yin:2023dbw}.

The basic axioms of the AO scheme are simple.
(\emph{a}) There is an effective charge, $\alpha_{1\ell}(k^2)$, in the sense of Refs.\,\cite{Grunberg:1980ja, Grunberg:1982fw}, reviewed in Ref.\,\cite{Deur:2023dzc}, that, when used to integrate the leading-order perturbative DGLAP equations \cite{Dokshitzer:1977sg, Gribov:1971zn, Lipatov:1974qm, Altarelli:1977zs}, defines an evolution scheme for every parton distribution function (DF) that is all-orders exact.
The form of $\alpha_{1\ell}(k^2)$ is largely irrelevant.  Nevertheless, the process-independent effective charge defined and computed in Refs.\,\cite{Binosi:2014aea, Binosi:2016nme, Cui:2019dwv} has all requisite properties.
(\emph{b}) There is a scale, $\zeta_{\cal H}<m_p$, at which all properties of a given hadron are carried by its valence degrees-of-freedom.  At this scale, DFs associated with glue and sea quarks are zero.  Nonzero values for glue and sea DFs are obtained via AO evolution to $\zeta > \zeta_{\cal H}$.

The AO approach extends DGLAP evolution onto QCD's nonperturbative domain.
It has been used successfully in many applications, \emph{e.g}.,
delivering unified predictions for pion fragmentation functions \cite{Xing:2023pms} and all pion, kaon, and proton DFs \cite{Cui:2020tdf, Chang:2022jri, Lu:2022cjx, Cheng:2023kmt, Yu:2024qsd},
a viable species separation of nucleon gravitational form factors \cite{Yao:2024ixu},
and insights into quark and gluon angular momentum contributions to the proton spin \cite{Yu:2024ovn}.

\section{Gap and Bethe-Salpeter Equations}
\label{ELequations}
In order to calculate the valence DF via Eq.\,\eqref{qFULL}, one needs the dressed quark propagator and pseudoscalar meson Bethe-Salpeter amplitude.
The quark propagator may be obtained from the QCD gap equation \cite{Roberts:1994dr}.
Thereafter, it can used to complete the kernel of the homogeneous Bethe-Salpeter equation and proceed to its solution, \emph{i.e}., the required Bethe-Salpeter amplitude.
Owing to the axialvector Ward-Green-Taka\-hashi identity, the kernels of both these equations are intimately connected \cite{Munczek:1994zz, Bender:1996bb}.
This deep link is readily expressed in modern applications of continuum Schwinger function methods (CSMs) because once the effective charge and gluon-quark vertex in the gap equation are known, it is straightforward to calculate a symmetry-consistent Bethe-Salpeter kernel \cite{Qin:2020jig, Xu:2022kng}.

\begin{table*}[t]
\caption{\label{TabMeson}
Masses and leptonic decay constants (in GeV) of selected mesons as obtained using the gap equation kernel specified by Eqs.\,\eqref{Amunu}, \eqref{vertex}.
Empirical values drawn from Ref.\,\cite[PDG]{ParticleDataGroup:2024cfk}.
Ground state pion, $\pi_0$, and first radial excitation, $\pi_1$, are listed.
The $b_1$ leptonic decay constant vanishes identically in the isospin symmetry limit, which we employ throughout.
A blank entry means no empirical value is available.
 }
\begin{center}
\begin{tabular*}
{\hsize}
{
l@{\extracolsep{0ptplus1fil}}
|l@{\extracolsep{0ptplus1fil}}
l@{\extracolsep{0ptplus1fil}}
l@{\extracolsep{0ptplus1fil}}
l@{\extracolsep{0ptplus1fil}}
l@{\extracolsep{0ptplus1fil}}
l@{\extracolsep{0ptplus1fil}}
l@{\extracolsep{0ptplus1fil}}}\hline\hline
meson $\ $ & $\pi_0\ $ & $\rho\ $ & $K\ $ & $K^\ast\ $ & $a_1\ $ & $b_1\ $ & $\pi_1\ $ \\\hline
mass calculated $\ $ & $0.14\ $ & $0.77\ $ & $0.494\ $ & $0.868\ $ & $1.20\ $ & $1.17\ $ & $1.24\ $ \\
$\rule{2.2em}{0ex}$ empirical $\ $ & $0.14\ $ & $0.77\ $ & $0.494\ $ & $0.892\ $ & $1.23(4)\ $ & $1.230(3)\ $ & $1.3(1)\ $ \\\hline
decay constant calculated $\ $ & $0.098\ $ & $0.159\ $ & $0.110\ $ & $0.178\ $ & $0.136\ $ & $0\ $ & $0.008\ $ \\
$\rule{6.5em}{0ex}$ empirical $\ $ & $0.092\ $ & $0.153(1)\ $ & $0.110(1)\ $ & $0.159(1)\ $ & $\ $ & $\ $ & $\ $
 \\\hline\hline
\end{tabular*}
\end{center}
\end{table*}

Analyses of QCD's gauge sector \cite{Binosi:2014aea, Binosi:2016nme, Cui:2019dwv} have enabled calculation of the QCD analogue of the Gell-Mann--Low effective charge that characterises QED \cite{GellMann:1954fq}.
This process-independent (PI) QCD charge is described in Ref.\,\cite{Brodsky:2024zev} and placed in context with pro\-cess-dependent charges in Ref.\,\cite{Deur:2023dzc}.  Herein, we use a representation of this charge, developed via bot\-tom-up analyses of hadron observables \cite{Binosi:2014aea}, to write the gap equation in the following form:
{\allowdisplaybreaks
\begin{subequations}
\label{EqGap}
\begin{align}
S^{-1}(k) &= i\gamma\cdot k + m + \Sigma(k) \,,\\
\Sigma(k) & =\int\frac{d^4 q}{(2\pi)^4}
4 \pi {\mathpzc A}_{\mu\nu}(l)
\gamma_\mu\frac{\lambda^{a}}{2} S(q) \Gamma_\nu(q,k)\frac{\lambda^{a}}{2}\,,
\end{align}
\end{subequations}
where we have suppressed all dependence on the renormalisation scale, $\zeta$.  (A discussion of our renormalisation procedure can be found, \emph{e.g}., in Ref.\, \cite{Chang:2008ec}: it is straightforward, mechanical, and symmetry-preserving.)
In\linebreak Eq.\,\eqref{EqGap},
$l=k-q$;
$m$ is the quark current-mass;\linebreak
$\{\tfrac{1}{2}\lambda^a|a=1,\ldots,8\}$ are the generators of SU$(3)$-colour in the fundamental representation;
${\mathpzc A}_{\mu\nu}$ is the effective vector-boson exchange interaction;
and $\Gamma_\nu$ is the relevant dres\-sed gluon-quark vertex.
The solution of Eq.\,\eqref{EqGap} is often written
$S(k) = 1/[i\gamma\cdot k\,A(k^2) + B(k^2)]$, a fact we used when writing Eq.\,\eqref{gtrE}.
}

Our approximation to the PI charge is introduced by writing \cite{Qin:2011dd, Binosi:2014aea}:
\begin{subequations}
\label{Amunu}
\begin{align}
{\mathpzc A}_{\mu\nu}(l)
& = T_{\mu\nu}(l) \tilde{\mathpzc A}(y=l^2) \,, \\
\tilde{\mathpzc A}(y) &
= 2\pi \frac{D}{\omega^4} {\rm e}^{-\tfrac{y}{\omega^2}}
+ \frac{2\pi \gamma_m \mathcal{F}(y)}{\ln\big[ \tau+(1+y/\Lambda_{\rm QCD}^2)^2 \big]}\,,
\end{align}
\end{subequations}
where $\gamma_m=12/25$, $\Lambda_{\rm QCD} = 0.234\,$GeV, $\tau={\rm e}^2-1$, and ${\cal F}(y) = \{1 - \exp(-y/\Lambda_{\mathpzc I}^2)\}/y$, $\Lambda_{\mathpzc I}=1\,$GeV.
This function is not a pointwise approximation to the interaction at nonperturbative momenta, $l^2 < (2 m_p)^2$.  Instead, it was developed to effectively reproduce global properties of the integral on this domain when combined with a curtailed expression for the gluon-quark vertex.
The tensor structure in Eq.\,\eqref{Amunu}, specified by $k^2 T_{\mu\nu}(l)=l^2\delta_{\mu\nu} - l_\mu l_\nu$, corresponds to Landau gauge, used because it is a fixed point of the renormalisation group.

Regarding the gluon-quark vertex, we follow Ref.\,\cite{Xu:2022kng} in using
\begin{equation}
\label{vertex}
  \Gamma_\nu(q,k) = \gamma_\nu + \eta \kappa(l^2) l_\alpha \sigma_{\alpha\nu}\,,
\end{equation}
$\kappa(l^2) = (1/\omega) \exp(-l^2/\omega^2)$.
(A possible overall multiplicative factor is implicitly absorbed into $\tilde{\mathpzc A}$.)
With $\eta=0$, one recovers the rainbow-ladder (RL) truncation, which is leading-order in the systematic, symmetry-preserving CSM approximation scheme introduced in Refs.\,\cite{Munczek:1994zz, Bender:1996bb}.  For reasons that have long been understood \cite{Fischer:2009jm, Chang:2009zb, Qin:2011xq, Chang:2011ei}, RL truncation is reliable for ground state hadrons with little rest-frame orbital angular momentum, but it requires improvement when dealing with excited states.
Such improvement is achieved with $\eta >0$ because the second term in Eq.\,\eqref{vertex} introduces a dressed-quark anomalous chromomagnetic moment (ACM),\linebreak which is known to appear as a corollary of EHM \cite{Chang:2010hb}.
This simple correction is capable of remedying the defects of RL truncation when treating the spectrum and leptonic decays of mesons built from light and/or\linebreak strange quarks \cite{Xu:2022kng}.
Widespread use has demonstrated \cite{Ding:2022ows} that interactions of the type in Eq.\,\eqref{Amunu} can serve to unify the properties of many systems.

We solve the gap and Bethe-Salpeter equations using what are now standard algorithms \cite{Maris:1997tm, Maris:2005tt, Krassnigg:2009gd}.
Contemporary studies employ $\omega = 0.8\,$GeV \cite{Xu:2022kng}.
Then, with $D = (0.68\,{\rm GeV})^2$,
$\eta = 1.1$,
and renormalisation point invariant quark current masses
\begin{equation}
\hat m_u = \hat m_d = 4.07\,{\rm MeV},
\quad
\hat m_s = 110\,{\rm MeV},
\end{equation}
which correspond to one-loop masses at $\zeta=2\,$GeV of $m_u=2.82\,$MeV, $m_s = 76.2\,$MeV -- in fair agreement with empirical estimates \cite[PDG]{ParticleDataGroup:2024cfk}, one obtains a good description of the lighter-quark meson spectrum.  An illustrative excerpt from the full array of results is presented in Table~\ref{TabMeson}.  It is worth highlighting that, compared with results obtained using RL truncation, inclusion of the ACM in Eq.\,\eqref{vertex} leads to a roughly $4.5$-fold reduction in the mean absolute relative difference between calculated and empirical masses \cite{Xu:2022kng}.
%

\section{Predictions for Valence DF Moments}
\label{SecPred}
At this point, with quark propagator and meson Bethe-Salpeter amplitude in hand, one can return to Eq.\,\eqref{qFULL}.  As explained elsewhere \cite{Chang:2013pq, Ding:2019qlr, Ding:2019lwe}, given that standard treatments of hadron bound state problems are formulated in Euclidean space \cite[Sect.\,1]{Ding:2022ows}, direct calculation of the $x$-dependence of the valence quark DF using numerical input for the integrand functions is difficult owing to the light-front projection.  We therefore elect to calculate a collection of lower-order moments:
{\allowdisplaybreaks
\begin{subequations}
\label{DFmoments}
\begin{align}
\langle x^m & \rangle_{{\mathpzc q}^\pi}^{\zeta_{\cal H}}
= \int dx x^m {\mathpzc q}^\pi(x;\zeta_{\cal H}) \\
& =
 \frac{N_c}{n\cdot P} {\rm tr}\! \int \frac{d^4k}{(2\pi)^4}\!
 \left[ \frac{n\cdot k_\eta}{n\cdot P} \right]^m
 \Gamma_\pi^P(k_{\bar\eta\eta};\zeta_{\cal H})\, S(k_{\bar\eta};\zeta_{\cal H})
\nonumber \\
& \times \{n\cdot\frac{\partial}{\partial {k_\eta}} \left[ \Gamma_\pi^{-P}(k_{\eta\bar\eta};\zeta_{\cal H}) S(k_\eta;\zeta_{\cal H}) \right]\}\,,
\end{align}
\end{subequations}
using which the DF can subsequently be reconstructed.
}

In principle, the angle integration in Eq.\,\eqref{DFmoments} ensures that the result for each moment is finite.
However, as $m$ increases, numerical accuracy becomes an issue.
The mass of the $\pi_1$ is such that $n\cdot P$ does not amplify integrand oscillations; consequently, standard algorithms provide reliable access to all $m\leq 6$ moments.
On the other hand, with $n\cdot P$ being small for the $\pi_0$, a little more sophistication is required in this case.
We proceed by exploiting the fact that the $\pi_0$ Bethe-Salpeter amplitude does not depend strongly on the angle $\theta$ defined by $k\cdot P$, so this dependence may be neglected for sufficiently large $k^2$.
Namely, we write
\begin{align}
  \Gamma_\pi^P(k_{\bar\eta\eta};\zeta_{\cal H}) & \to
  H(\Lambda_\theta^2 - q^2) \Gamma_\pi^P(k_{\bar\eta\eta};\zeta_{\cal H}) \nonumber \\
 & \quad + H(q^2- \Lambda_\theta^2) \left. \Gamma_\pi^P(k_{\bar\eta\eta};\zeta_{\cal H})\right|_{\theta=0}\,,
\end{align}
where $H(t)$ is the Heaviside step function;
calculate each moment as a function of $\Lambda_\theta$;
then define the moment to be the value obtained when the result for the integration becomes independent of $\Lambda_\theta$.
In this way, with $\Lambda_\theta = 4\,$GeV, one obtains stable results for $m\leq 5$.
This scheme is easier to implement and at least as reliable as the alternative employed in Ref.\,\cite{Ding:2019lwe} -- see Eq.\,(30) therein.
Notably, since pion-like hadron-scale valence quark DFs are known to be symmetric functions, Eq.\,\eqref{valencemom}, any given odd moment is completely determined by the lower order even moments -- see, \emph{e.g}., Ref.\,\cite[Eq.\,(29)]{Ding:2019lwe}.
Confirming these recursion relations provides a valuable check on the accuracy of the moment determinations.

Using the set of reliable moments, we follow Ref.\,\cite{Ding:2019lwe} and extend the range to $m\leq 10$ using the Schlessinger point method (SPM) \cite{Schlessinger:1966zz, PhysRev.167.1411, Tripolt:2016cya, Cui:2022fyr}.
In this way, we construct an analytic function, $M(z)$, which precisely reproduces the known moments, respecting the recursion relations, and define the higher-order moments to be the values of $M(z)$ at those integer values.
In nontrivial test cases, $M(z)$ returns moments out to $m=10$ whose relative error is $< 0.1$\% in magnitude.

\begin{table}[t]
\caption{\label{moments}
Hadron scale valence DF moments of the identified system.
The superscript ``RL'' means calculated in RL truncation.
Column $\pi_1^{(a)}$ -- all $m\leq 5$ moments are truncated to three significant figures; and
column $\pi_1^{(b)}$ -- $m=3, 5$ moments are the precise results obtained from the lower-$m$ even moments when using the applicable recursion relation \cite[Eq.\,(29)]{Ding:2019lwe}.
The moments of the scale free DF, Eq.\,\eqref{sfDF}, are
$\langle x^m\rangle_{\rm sf} = 60/[(3+m)(4+m)(5+m)]$.
In all cases, the $m=0$ moment is unity (baryon number conservation) and $m=1$ is $1/2$ (momentum conservation at $\zeta_{\cal H}$).
 }
\begin{center}
\begin{tabular*}
{\hsize}
{
l@{\extracolsep{0ptplus1fil}}
|l@{\extracolsep{0ptplus1fil}}
l@{\extracolsep{0ptplus1fil}}
l@{\extracolsep{0ptplus1fil}}
l@{\extracolsep{0ptplus1fil}}
l@{\extracolsep{0ptplus1fil}}}\hline
moment $m\ $ & $\pi_0\ $ & $\pi_0^{\rm RL}\ $ & $\pi_1^{(a)}\ $  & $\pi_1^{(b)}\ $ & scale free \\\hline
2 & 0.297 & 0.302 & 0.275 & 0.275 & 0.286 \\
3 & 0.195 & 0.203 & 0.162 & 0.1625 & 0.179 \\
4 & 0.137 & 0.146 & 0.104 & 0.104 & 0.119 \\
5 & 0.102 & 0.109 & 0.0725 & 0.0725 & 0.0833\\
6 & 0.0788 & 0.0848 & 0.0541 & 0.0548 & 0.0606\\
7 & 0.0628 & 0.0673 & 0.0426 & 0.0444 & 0.0455\\
8 & 0.0513 & 0.0544 & 0.0349 & 0.0379 & 0.0350\\
9 & 0.0428 & 0.0446 & 0.0296 & 0.0337 & 0.0275\\
10 & 0.0363 & 0.0370 & 0.0256 & 0.0307 & 0.0220
\\\hline
\end{tabular*}
\end{center}
\end{table}

The valence DF moments of the ground-state pion and its first radial excitation are listed in Table~\ref{moments}, where\-in they are compared with moments of the scale-free DF:
\begin{equation}
\label{sfDF}
{\mathpzc q}^{\rm sf}(x) = 30 x^2 (1-x)^2 .
\end{equation}
Results obtained in the Ref.\,\cite{Ding:2019lwe} RL truncation treatment of the ground-state pion are also listed in Table~\ref{moments}.
Plainly, they are not materially different from those calculated herein using the EHM-improved kernels described in Sect.\,\ref{ELequations}.
Owing to Eq.\,\eqref{BSAmp}, a corollary of the axialvector Ward-Green-Takahashi identity, this was to be expected.

Two sets of $\pi_1$ moments are listed in Table~\ref{moments}.
They correspond to SPM extrapolations based on moments $0\leq m\leq 5$ in their respective columns.
In the $\pi_1^{(a)}$ column, all source moments were truncated to three significant figures.
On the other hand, the $\pi_1^{(b)}$ $m=3$ moment is the precise result obtained from the $m=0,2$ moments when using the applicable recursion relation \cite[Eq.\,(29b)]{Ding:2019lwe}.
The modest impact on the extrapolation of the additional significant figure is apparent.
We use both $\pi_1$ columns to arrive at a DF for this radial excitation with an indicative uncertainty.
Such a procedure is unnecessary for the ground state because, as will become apparent, the DF for this system is much simpler.

We have calculated RL moments for the $\pi_1$.  However, as noted above, RL truncation is not reliable for excited states.
In this case, its use cannot guarantee
$(\langle x \rangle_{{\mathpzc q}^{\pi_1}}^{\zeta_{\cal H}})^2
< \langle x^2 \rangle_{{\mathpzc q}^{\pi_1}}^{\zeta_{\cal H}}$, a condition that must be satisfied by moments of a non-negative DF.  (This is an expression of the Cauchy-Schwarz inequality.)
The problem originates in RL truncation delivering a dressed quark mass function that is too large at infrared momenta, $k^2 \lesssim (2 m_p)^2$, and a Bethe-Salpeter amplitude in which the magnitudes of the Poincar\'e-covariant coefficient functions damp too slowly with increasing $k^2$.  Namely, RL delivers a poor structural picture of pion radial excitations.
For this reason, Table~\ref{moments} does not include RL moments for the $\pi_1$.

Before continuing to a discussion of pointwise DF reconstruction, we note that the moments in Table~\ref{moments} send strong signals regarding the $x$-dependence that should be anticipated.
The $\pi_0$ moments are uniformly larger than those of the scale-free DF; hence, with respect to ${\mathpzc q}^{\rm sf}$, the $\pi_0$ DF should exhibit significant dilation and, therefore, a lower peak height.
This has long been known \cite{Ding:2019qlr, Ding:2019lwe}.
On the other hand, the low-$m$ $\pi_1$ moments are smaller than those of ${\mathpzc q}^{\rm sf}$; so, the $\pi_1$ DF should be narrower and taller on $x\simeq 1/2$.  Yet, the large-$m$ $\pi_1$ moments are greater than those of ${\mathpzc q}^{\rm sf}$.
Consequently, the $\pi_1$ DF should exhibit greater support on the endpoint domains.
These remarks suggest an interesting structural picture of the $\pi_1$.

\begin{table}[t]
\caption{\label{leastsquares}
DF reconstruction parameters for using in Eqs.\,\eqref{CSMDF}, \eqref{CSMDFradial}.
In each case, the first row reports the mean absolute relative difference (in \%) between the relevant $m\geq 3$ moments in Table~\ref{moments} and the values calculated using the reconstructed DF.
 }
\begin{center}
\begin{tabular*}
{\hsize}
{
l@{\extracolsep{0ptplus1fil}}
|l@{\extracolsep{0ptplus1fil}}
l@{\extracolsep{0ptplus1fil}}
l@{\extracolsep{0ptplus1fil}}
l@{\extracolsep{0ptplus1fil}}}\hline
        & $\ \pi_0\ $ & $\ \pi_0^{\rm RL}\ $ & $\quad \pi_1^{(a)}\ $  & $\ \pi_1^{(b)}\ $  \\\hline
 $\overline{\rm ard}\ $ & 1.7(8) & 4.1(1.4) & $\phantom{-}1.4(1.0)\ $& 5.0(2.8) \\
$\rho\ $ & 0.0750 & 0.0613 & $\phantom{-}0.743\ $ & 0.663 \\
$a_2\ $ & 0 & 0  & $\phantom{-}0.116\ $ & 0.191 \\
$a_4\ $ & 0 & 0 & $\phantom{-}0.441\ $ & 0.487 \\
$a_6\ $ & 0 & 0 & $-0.00300\ $ & 0.0684 \\
\hline
\end{tabular*}
\end{center}
\end{table}

\section{DF Reconstruction}
\label{DFRecon}
In studies of pseudoscalar meson valence-quark DFs that incorporate the ultraviolet behaviour of the pion wave function prescribed by QCD, one finds \cite{Cui:2021mom}:
\begin{equation}
\label{endpointzH}
{\mathpzc q}^\pi(x;\zeta_{\cal H}) \stackrel{x\simeq 1}{\sim}  (1-x)^2,
\end{equation}
with analogous behaviour on $x\simeq 0$ owing to Eq.\,\eqref{valencemoma}.  These features can be exploited to simplify reconstruction of the valence DF of systems built from mass-degenerate valence degrees of freedom.  Indeed, for \linebreak ground states of such systems the following function is efficacious \cite{Cui:2022bxn, Lu:2023yna}:
\begin{equation}
{\mathpzc q}^{\pi_0}(x;\zeta_{\cal H}) =
{\mathpzc n}_{\pi} \ln[ 1+x^2(1-x)^2/\rho_\pi^2] =: \tilde {\mathpzc q}(x;\zeta_{\cal H})  \,,
\label{CSMDF}
\end{equation}
with $\rho_\pi$ a reconstruction parameter and ${\mathpzc n}_\pi$ a constant that ensures unit normalisation.
The function in Eq.\,\eqref{CSMDF} has the merits that it is flexible enough to simultaneously express both the dilation that EHM is known to produce in the valence quark DF \cite{Lu:2022cjx} and endpoint behaviour matching QCD expectations, Eq.\,\eqref{endpointzH}.

Returning to Table~\ref{moments}, it appears that somewhat more flexibility is required in order to describe the $\pi_1$ DF.  We use
\begin{equation}
{\mathpzc q}^{\pi_1}(x;\zeta_{\cal H}) =
\tilde {\mathpzc q}(x;\zeta_{\cal H}) \big[1+\sum_{i=1,2,3} a_{2i} C_{2i}^{3/2}(1-2x)\big]\,,
\label{CSMDFradial}
\end{equation}
where $C_n^{3/2}$ is a Gegenbauer polynomial of order $3/2$.
Here, $\rho_{\pi_1}$, $\{a_{2i}\}$ are the reconstruction parameters.

\begin{figure}[t]
\centerline{%
\includegraphics[clip, width=0.45\textwidth]{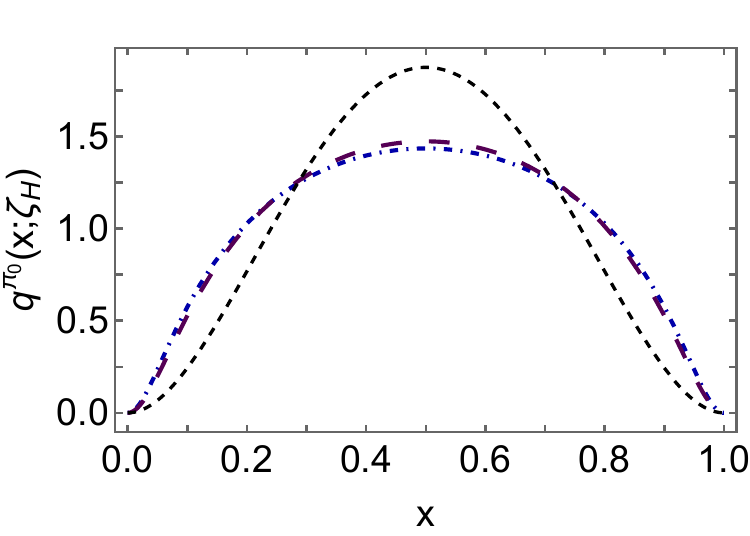}}
\caption{\label{FigDF0}
Hadron-scale valence DFs in the pion ground state:
long-dashed purple curve -- result obtained using EHM-improved bound-state kernels, Sect.\,\ref{ELequations};
dot-dashed blue curve -- RL result.
The scale-free DF, Eq.\,\eqref{sfDF}, is drawn for comparison -- dashed black curve.
}
\end{figure}

To begin, consider the first two columns in Table~\ref{moments}.
Since only the even moments are independent, then for both the EHM-improved and RL treatments of the $\pi_0$ we determine $\rho_\pi$ by minimising the mean-square difference between the even source moments and the values calculated using Eq.\,\eqref{CSMDF}.
The results are listed in Table~\ref{leastsquares}, with the associated reconstructed DFs depicted in Fig.\,\ref{FigDF0}.
Evidently, for the ground state, there is little to distinguish between the DFs produced by the sophisticated and simplest kernels, \emph{viz}.\ largely irrespective of the kernels, in a clear expression of EHM, ${\mathpzc q}^{\pi_0}(x;\zeta_{\cal H})$ is much dilated with respect to the scale-free profile \cite{Ding:2019qlr, Ding:2019lwe}.
Furthermore, the DFs drawn in Fig.\,\ref{FigDF0} are concordant with the relation
\begin{equation}
{\mathpzc q}^{\pi_0}(x;\zeta_{\cal H}) \propto |\varphi_{\pi_0}(x;\zeta_{\cal H})|^2,
\label{DFDA2}
\end{equation}
highlighted and discussed elsewhere \cite{Cui:2020tdf, Raya:2021zrz, Roberts:2021nhw}.

\begin{figure}[t]
\centerline{%
\includegraphics[clip, width=0.45\textwidth]{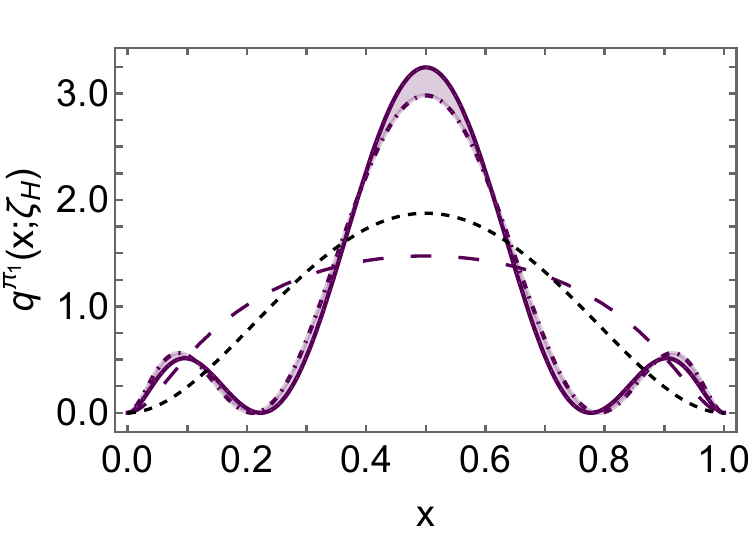}}
\caption{\label{FigDF1}
Hadron-scale valence DFs in the lightest pion radial excitation:
solid purple curve -- result obtained using $\pi_1^{(a)}$ column of moments from Table~\ref{moments};
dot-dashed purple curve -- obtained using $\pi_1^{(b)}$ column of moments.
The $\pi_0$ (long-dashed purple) and scale-free (dashed black) DFs from Fig.\,\ref{FigDF0} are drawn for comparison.
}
\end{figure}

Having established that our methods reproduce \linebreak known results, we can now address one of our principal goals, \emph{i.e}., reconstruction of the $\pi_1$ valence DF.
Working with the even source moments in the third and fourth columns of Table~\ref{moments}, we determine the Eq.\,\eqref{CSMDFradial} reconstruction parameters via a least-squares minimisation procedure analogous to that used for the ground state.
The only difference is that we also impose a DF non-negativity constraint on the parameter search because unfettered reliance on an expansion in Gegenbauer polynomials can deliver large-magnitude oscillations.
This procedure yields the reconstruction parameters in the third and fourth columns of Table~\ref{leastsquares}, with the associated DFs drawn in Fig.\,\ref{FigDF1}.
Recall that for the radial excitation, only the EHM-improved kernels, discussed in Sect.\,\ref{ELequations}, provide sound results.

Regarding Fig.\,\ref{FigDF1}, it is plain that the two sets of Mellin moments yield qualitatively identical $\pi_1$ DFs.  The small quantitative differences between the two reconstructions are a fair expression of the uncertainty in the true moments.

The most striking feature of the $\pi_1$ DF in Fig.\,\ref{FigDF1} is its three-peak structure.
As anticipated in closing Sect.\,\ref{SecPred}, the $x=1/2$ peak is much narrower and taller than the analogous feature of the scale-free DF -- this explains the relative size of the low-$m$ moments obtained from these two DFs.
Then the competing requirement that larger-$m$ $\pi_1$ moments are greater than those of ${\mathpzc q}^{\rm sf}$ is met by the appearance of the $x\sim 0.2, 0.8$ minima and subsequent secondary peaks, which enable this DF to possess support on the endpoint domains that is materially greater than that owned by ${\mathpzc q}^{\rm sf}$.
Finally, each of the two zero-value minima of the $\pi_1$ valence DF is located in the vicinity of one of the two zeroes in the $\pi_1$ DA -- see Fig.\,\ref{FigDA1}.
This is a signal that Eq.\,\eqref{DFDA2} is, at least qualitatively, expressed in $\pi_1$ structure.

In all these things, the $\pi_1$ valence DF is exhibiting features analogous to the modulus-squared of the quantum mechanics momentum-space wave function of a bound-state's first radial excitation.
Importantly, the modulus-squared of a light-front wave function possess a probability interpretation.
Consider first the ground state: ${\mathpzc q}^{\pi_0}(x;\zeta_{\cal H})$ has a single peak at $x=1/2$ and vanishes at the endpoints.  Given the above remarks, then these features may be interpreted as indicating that the momentum-fraction probability density favours zero relative momentum ($x=1/2$) within the ground state and disfavours the extreme case that one dressed valence degree of freedom carries all or none of the momentum.
Then, the meaning of the three-peak structure of\linebreak ${\mathpzc q}^{\pi_1}(x;\zeta_{\cal H})$ is clear by analogy.
Namely, in the radial excitation, too, zero relative momentum between the dressed valence degrees of freedom is favoured and the extreme all-or-none case disfavoured.
Furthermore, momentum fractions $x \approx 0.2,0.8$ are equally disfavoured, with that support transferred to heightened probability density in the domains centred on $x\approx 0.1,0.9$ and $x=1/2$.

\section{DFs beyond the Hadron Scale}
\label{SecAOE}
Hitherto, we have focused on determining the $\pi_{0,1}$ valence DFs at $\zeta_{\cal H}$, whereat valence degrees of freedom carry all properties of a given hadron and, consequently, glue and sea DFs vanish.
Naturally, all glue and sea partonic Fock space components are sublimated into the valence degrees of freedom at $\zeta_{\cal H}$ because this is the meaning and impact of dressing, as may be seen by studying the diagrammatic content of the quantum field equations satisfied by all elements in any symmetry-preserving CSM analysis of a given observable.   As the resolving scale increases beyond $\zeta_{\cal H}$, however, the dres\-sed valence degrees of freedom begin to shed their clothing, so that, \emph{e.g}., glue and sea DFs become nonzero.  This physics is described by the DGLAP evolution equations \cite{Dokshitzer:1977sg, Gribov:1971zn, Lipatov:1974qm, Altarelli:1977zs} and we express these processes by employing the AO evolution scheme sketched in closing Sect.\,\ref{secDFformula}.

As noted above, in principle, it is not necessary to specify the hadron scale when employing AO evolution.  Nevertheless, if one chooses a particular effective charge, then the value becomes known.
The process-independent effective charge calculated in Ref.\,\cite{Cui:2019dwv} has proved useful.
This charge defines a screening mass, whose value is a natural choice for the hadron-scale: $\zeta_{\cal H}=0.331(2)\,$GeV.
Analysis of results from lat\-tice-regularised QCD (lQCD) relating to the $\pi_0$ valence quark DF yields a consistent value \cite{Lu:2023yna}: $\zeta_{\cal H} = 0.350(44)\,{\rm GeV}$.

\begin{figure}[t]
\centerline{%
\includegraphics[clip, width=0.45\textwidth]{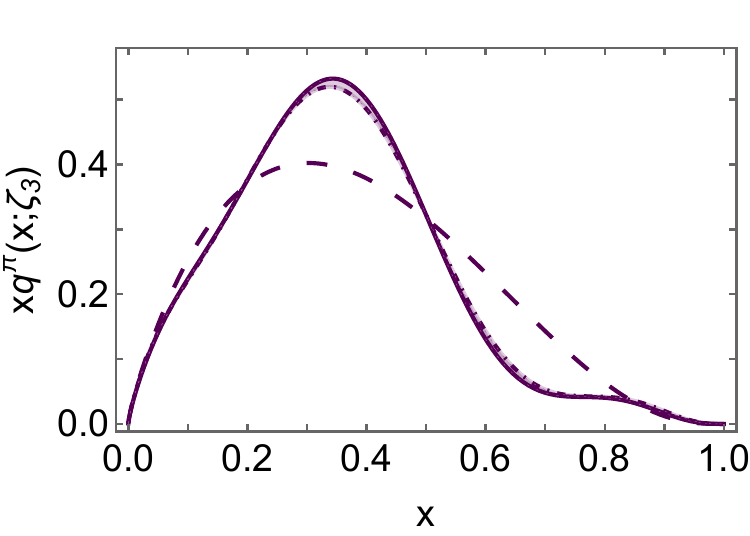}}
\caption{\label{FigDFz3}
Valence quark DFs in the pion and its first radial excitation, at resolving scale $\zeta_3=3.2\,$GeV:
long-dashed -- ground state pion, $\pi_0$;
dot-dashed -- obtained using $\pi_1^{(b)}$ column of moments in Table~\ref{moments};
solid curve -- result obtained using $\pi_1^{(a)}$ column of moments.
}
\end{figure}

In Ref.\,\cite{Gao:2021hvs}, using lQCD, an exploratory effort was made to identify differences between $\pi_{0,1}$ valence quark structure at the resolving scale $\zeta_3=3.2\,$GeV.
We use this scale hereafter in comparing valence, glue and sea DFs in these systems.
Before proceeding toward that goal, however, we remark that the lQCD study reported results for the lowest four nontrivial Mellin moments of the valence DFs, $m=1,2,3,4$: in all cases, $\langle x^m\rangle_{{\mathpzc q}^{\pi_1}}^{\zeta_3} > \langle x^m\rangle_{{\mathpzc q}^{\pi_0}}^{\zeta_3}$.
This ordering is incompatible with the results reported in Table~\ref{moments}.
Moreover, the ordering of valence DF moments is not changed by DGLAP evolution,\linebreak whether perturbative or using the AO scheme.
In our view, therefore, it is necessary for additional lQCD analyses to be completed before it becomes possible to judge the potential ability of the lattice approach to deliver reliable information on excited-state DFs.

In Fig.\,\ref{FigDFz3}, we compare $\pi_{0,1}$ valence quark DFs after AO evolution to $\zeta_3$.
The hadron-scale differences between these DFs are reduced by evolution.
For instance, the absolute minima in the $\pi_1$ DF have become diffuse regions containing an inflection point.
Nevertheless, in principle, the $\pi_{0,1}$ differences remain empirically discernible to precise structure function measurements combined with sound phenomenological analyses.
Of course, since $\pi_1$ targets, real or virtual, are unlikely to ever be achievable, $\pi_1$ structure function measurements will remain improbable.
Nevertheless, our predictions can serve as valuable benchmarks for other theory attempts to expose structural features of the pion and its radial excitations.
For that reason, we list the first four nontrivial moments of each valence DF in Table~\ref{momentsz3}.
As indicated above, evolution does not alter the relative sizes of ground-state vs.\ radial excitation moments.  The uncertainty on the hadron-scale $\pi_1$ valence DF -- highlighted in Fig.\,\ref{FigDF1} -- is reflected in slight differences between the evolved moments in columns~2 and 3.

\begin{table}[t]
\caption{\label{momentsz3}
Selected Mellin moments of $\pi_{0,1}$ DFs -- valence ($\mathpzc V$), glue ($\mathpzc g$), and flavour separated sea ($\mathpzc S$) -- all determined at the scale $\zeta_3=3.2\,$GeV.  $G$-parity symmetry \cite{Lee:1956sw} entails that $u=d$ for valence and sea DFs in pion-like systems.
 }
\begin{center}
\begin{tabular*}
{\hsize}
{
l@{\extracolsep{0ptplus1fil}}
|l@{\extracolsep{0ptplus1fil}}
l@{\extracolsep{0ptplus1fil}}
l@{\extracolsep{0ptplus1fil}}
l@{\extracolsep{0ptplus1fil}}}\hline
    & $\pi_0\ $ & $\pi_1^{(a)}\ $ & $\pi_1^{(b)}\ $ \\ \hline
${\mathpzc V}\;\langle x \rangle_{{\mathpzc q}^\pi}^{\zeta_3}\ $ & 0.220 & 0.220 & 0.220 \\
$\phantom{{\mathpzc V}\;}\langle x^2 \rangle_{{\mathpzc q}^\pi}^{\zeta_3}\ $ & 0.0831 & 0.0772 & 0.0779 \\
$\phantom{{\mathpzc V}\;}\langle x^3 \rangle_{{\mathpzc q}^\pi}^{\zeta_3}\ $ & 0.0397 & 0.0334 & 0.0341 \\
$\phantom{{\mathpzc V}\;}\langle x^4 \rangle_{{\mathpzc q}^\pi}^{\zeta_3}\ $ & 0.0218 & 0.0167 & 0.0173 \\
\hline
${\mathpzc g}\;\langle x \rangle_{{\mathpzc g}^\pi}^{\zeta_3}\ $ & 0.434 & 0.434 & 0.434 \\
$\phantom{{\mathpzc g}\;}\langle x^2 \rangle_{{\mathpzc g}^\pi}^{\zeta_3}\ $ & 0.0346 & 0.0321 & 0.0324 \\\hline
${\mathpzc S}\;\langle x \rangle_{{\mathpzc u}^\pi}^{\zeta_3}\ $ & 0.0372 & 0.0372 & 0.0372 \\
$\phantom{{\mathpzc S}\;}\langle x \rangle_{{\mathpzc s}^\pi}^{\zeta_3}\ $ & 0.0308 & 0.0308 & 0.0308 \\
$\phantom{{\mathpzc S}\;}\langle x \rangle_{{\mathpzc c}^\pi}^{\zeta_3}\ $ & 0.0187 & 0.0187 & 0.0187 \\
$\phantom{{\mathpzc S}\;}\langle x^2 \rangle_{{\mathpzc u}^\pi}^{\zeta_3}\ $ & 0.00274 & 0.00254 & 0.00257 \\
$\phantom{{\mathpzc S}\;}\langle x^2 \rangle_{{\mathpzc s}^\pi}^{\zeta_3}\ $ & 0.00219 & 0.00204 & 0.00206 \\
$\phantom{{\mathpzc S}\;}\langle x^2 \rangle_{{\mathpzc c}^\pi}^{\zeta_3}\ $ & 0.00117 & 0.00109 & 0.00110 \\\hline
\end{tabular*}
\end{center}
\end{table}

\begin{figure}[t]
\vspace*{0ex}

\leftline{\hspace*{0.5em}{{\large\textsf{A}}}}
\vspace*{-4ex}
\centerline{\includegraphics[width=0.45\textwidth]{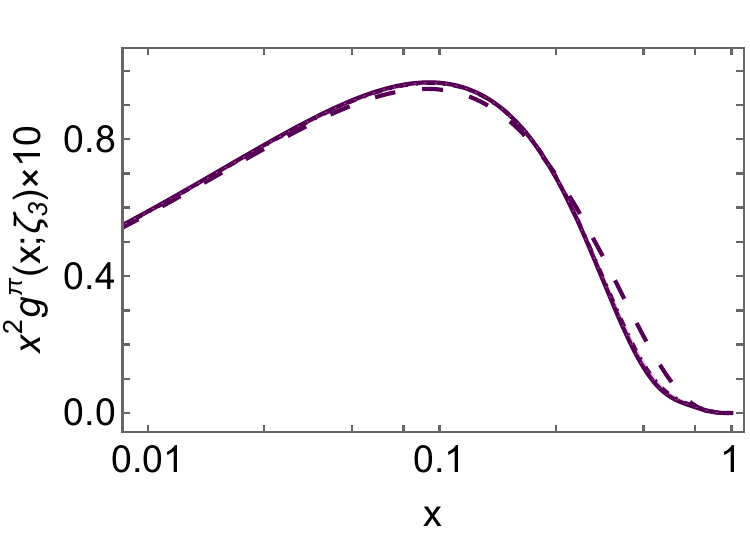}}
\vspace*{2ex}

\leftline{\hspace*{0.5em}{{\large\textsf{B}}}}
\vspace*{-3ex}
\centerline{\includegraphics[width=0.45\textwidth]{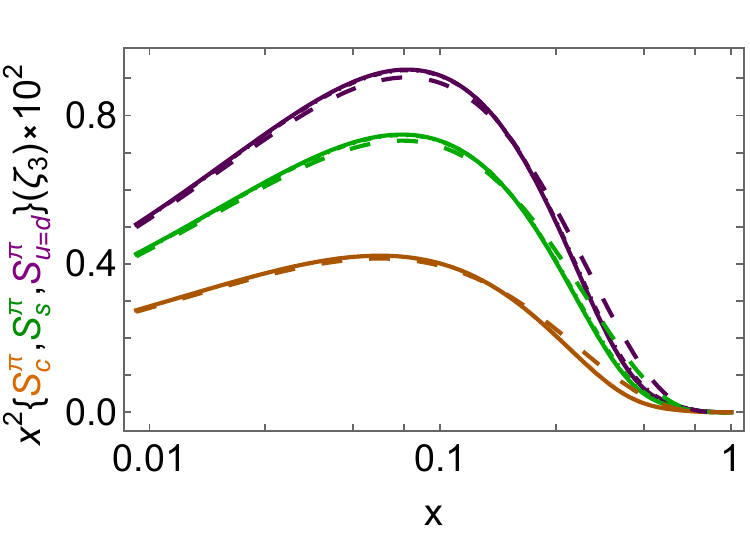}}

\caption{\label{FigDFgSz3}
\textsf{Panel A}. Glue DFs in the pion (long dashed) and its first radial excitation (solid curve).
\textsf{Panel B}. Analogous sea quark DFs: $\pi_0$ -- long dashed; $\pi_1$ -- solid.  Further, $c$ quark -- orange; $s$ quark -- green; $u=d$ -- purple.  These curves are also distinguished by their increase in peak magnitude, which is highlighted by the ordering of the $y$ axis labels.
(All curves calculated at $\zeta_3=3.2\,$GeV and, in both panels, results obtained using $\pi_1^{(a)}$ and $\pi_1^{(b)}$ valence DF reconstructions are practically indistinguishable.)
}
\end{figure}

Nonzero glue and sea DFs are exposed by evolution $\zeta_{\cal H}\to \zeta_3$.  We compare those for the ground-state pion and its first radial excitation in Fig.\,\ref{FigDFgSz3}.
Evidently, the ground-state DFs possess a small but noticeable amount of additional support on $0.25 \lesssim x\lesssim 0.75$.
Practically, however, even if $\pi_1$ DFs were accessible via structure function measurements, the differences are too small to be empirically discernible.
Notwithstanding that, they remain as reference points for other theoretical analyses.

It is worth highlighting a long known feature of DFs obtained using AO evolution.
Namely, with the same value of the hadron scale characterising every strong-interaction bound-state, then the momentum fraction carried by each species is the same in every hadron.
For instance, at a given resolving scale, glue carries the same light-front fraction of the hadron total momentum in both the pion and proton \cite{Lu:2022cjx}, and -- as we see now -- in the pion radial excitation, too.
Differences emerge when valence-quark current-mass effects are introduced into the splitting-functions of the evolution kernels \cite{Cui:2020tdf}.

\section{Summary}
\label{epilogue}
A symmetry-preserving approximation to the quantum field equations relevant in calculations of meson masses and interactions was used to deliver predictions for the distribution functions of all species -- valence, glue, and separated four-flavour sea -- in both the ground state pion, $\pi_0$, and its first radial excitation, $\pi_1$.
The kernels used in formulating the bound-state equations express nonperturbatively calculated im\-prove\-ments to the leading-order formulae that are usually employed; in particular, effects of a dressed-quark anomalous chromomagnetic moment.
Whilst this has little impact on ground-state structure, it is crucial to obtaining a plausible description of the structure of the radial excitation.

Working at the hadron scale, $\zeta_{\cal H}$, whereat valence degrees of freedom carry all properties of the target  hadron, we computed eleven Mellin moments of the $\pi_{0,1}$ valence DFs [Sect.\,\ref{SecPred}].  Labelling the moments by $m$, then, owing to baryon number and momentum conservation, the $m=0,1$ moments are identical in both systems.
However, for $m\geq 2$, each ground state moment is larger than its partner in the excited state.
In addition, while each $m\in [2,7]$ moment of the radial excitation DF is smaller than the matched moment in the scale-free DF, ${\mathpzc q}^{\rm sf}(x) = 30 x^2(1-x)^2$, they are greater for $m \geq 8$.

Working with this information, we developed pointwise reconstructions of the hadron-scale $\pi_{0,1}$ valence DFs [Sect.\,\ref{DFRecon}].
The ground-state DF thus obtained is in accord with existing results, an outcome which may be viewed as confirmation of our methods.
Hitherto, no information existed on hadron-scale $\pi_1$ DFs, and our result for the valence DF, ${\mathpzc q}^{\pi_1}(x;\zeta_{\cal H})$, has a curious but not wholly unexpected appearance [Fig.\,\ref{FigDF1}].
Namely, it possess three-peaks, with the central maximum partnered by secondary peaks on either side, each separated from the centre by a zero: the zeroes lie at $x\approx 0.2,0.8$ and the secondary peaks at $x\approx 0.1,0.9$.
In displaying these features, the form of ${\mathpzc q}^{\pi_1}(x;\zeta_{\cal H})$ is qualitatively consistent with the picture of this function as the modulus-squared of the meson's parton distribution amplitude.
The meaning follows therefrom; to wit, as in the ground state, zero relative momentum ($x \simeq 1/2$) between the dressed valence degrees of freedom is favoured and the extreme all-or-nothing cases ($x \simeq 0,1$) are strongly suppressed.
In addition, momentum fractions $x \simeq 0.2,0.8$ are also strongly suppressed, with the support relocated to domains centred on $x\approx 0.1,0.9$ and $x=1/2$.

Evolution from $\zeta_{\cal H}\to \zeta_3 = 3.2\,$GeV, a commonly used resolving scale for nonperturbative calculations, was accomplished using the all-orders scheme [Sect.\,\ref{secDFformula}], a well-tested extension of standard evolution onto QCD's nonperturbative domain.
At this higher scale, differences between the $\pi_{0,1}$ valence DFs are reduced, yet still significant on $0.1\lesssim x \lesssim 0.9$ [Fig.\,\ref{FigDFz3}].
Differences between glue and sea DFs in the ground state and its first radial excitation are far smaller.  In fact, they are practically indiscernible [Fig.\,\ref{FigDFgSz3}].

Our analysis has shown that, owing to constraints imposed by chiral symmetry and the pattern by which it is broken in quantum chromodynamics (QCD) -- the latter being an expression of emergent hadron mass, there are some noticeable differences between the structural properties of the pion ground state and its radial excitations.
Therefore, given that two distinct Hamiltonians which produce similar spectra, can often generate very different structural pictures of the bound states in those spectra, then other nonperturbative frameworks which aim at a reliable description of hadron properties should be tested against the predictions delivered herein.
Remedying any failures to, at least qualitatively, reproduce our results could then prove useful, \emph{e.g}., in refining models to better express key QCD symmetries and their corollaries, or improving algorithms or problem formulations in the numerical simulation of lattice-regularised QCD.


\begin{acknowledgements}
Work supported by:
National Natural Science Foundation of China (grant no.\ 12135007);
Spanish Ministry of Science, Innovation and Universities (MICIU grant no.\ PID2022-140440NB-C22);
%
%
Program d'Alembert,
Pro\-ject AST22$\_00001\_$X, funded by NextGenerationEU and ``Plan de Recuperaci\'on, Transformaci\'on y Resiliencia y Resiliencia'' (PRTR) from MICIU and Junta de Andaluc{\'{\i}}a;
and completed in part at Institut Pascal, Universit\'e Paris-Saclay, with the support of the program ``Investissements d'avenir'' ANR-11-IDEX-0003-01.
\end{acknowledgements}


\begin{thebibliography}{81}
\providecommand{\natexlab}[1]{#1}
\providecommand{\url}[1]{\texttt{#1}}
\providecommand{\urlprefix}{URL }
\expandafter\ifx\csname urlstyle\endcsname\relax
  \providecommand{\doi}[1]{doi:\discretionary{}{}{}#1}\else
  \providecommand{\doi}[1]{doi:\discretionary{}{}{}\begingroup
  \urlstyle{rm}\url{#1}\endgroup}\fi
\providecommand{\bibinfo}[2]{#2}

\bibitem[{Maris et~al.(1998)Maris, Roberts, and Tandy}]{Maris:1997hd}
\bibinfo{author}{P.~Maris}, \bibinfo{author}{C.~D. Roberts},
  \bibinfo{author}{P.~C. Tandy}, \bibinfo{title}{{Pion mass and decay
  constant}}, \bibinfo{journal}{Phys. Lett. B} \bibinfo{volume}{420}
  (\bibinfo{year}{1998}) \bibinfo{pages}{267--273}.

\bibitem[{Brodsky et~al.(2012)Brodsky, Roberts, Shrock, and
  Tandy}]{Brodsky:2012ku}
\bibinfo{author}{S.~J. Brodsky}, \bibinfo{author}{C.~D. Roberts},
  \bibinfo{author}{R.~Shrock}, \bibinfo{author}{P.~C. Tandy},
  \bibinfo{title}{{Confinement contains condensates}}, \bibinfo{journal}{Phys.
  Rev. C} \bibinfo{volume}{85} (\bibinfo{year}{2012}) \bibinfo{pages}{065202}.

\bibitem[{Qin et~al.(2014)Qin, Roberts, and Schmidt}]{Qin:2014vya}
\bibinfo{author}{S.-X. Qin}, \bibinfo{author}{C.~D. Roberts},
  \bibinfo{author}{S.~M. Schmidt}, \bibinfo{title}{{Ward-Green-Takahashi
  identities and the axial-vector vertex}}, \bibinfo{journal}{Phys. Lett. B}
  \bibinfo{volume}{733} (\bibinfo{year}{2014}) \bibinfo{pages}{202--208}.

\bibitem[{Horn and Roberts(2016)}]{Horn:2016rip}
\bibinfo{author}{T.~Horn}, \bibinfo{author}{C.~D. Roberts},
  \bibinfo{title}{{The pion: an enigma within the Standard Model}},
  \bibinfo{journal}{J. Phys. G.} \bibinfo{volume}{43} (\bibinfo{year}{2016})
  \bibinfo{pages}{073001}.

\bibitem[{Roberts et~al.(2021)Roberts, Richards, Horn, and
  Chang}]{Roberts:2021nhw}
\bibinfo{author}{C.~D. Roberts}, \bibinfo{author}{D.~G. Richards},
  \bibinfo{author}{T.~Horn}, \bibinfo{author}{L.~Chang},
  \bibinfo{title}{{Insights into the emergence of mass from studies of pion and
  kaon structure}}, \bibinfo{journal}{Prog. Part. Nucl. Phys.}
  \bibinfo{volume}{120} (\bibinfo{year}{2021}) \bibinfo{pages}{103883}.

\bibitem[{Binosi(2022)}]{Binosi:2022djx}
\bibinfo{author}{D.~Binosi}, \bibinfo{title}{{Emergent Hadron Mass in Strong
  Dynamics}}, \bibinfo{journal}{Few Body Syst.}
  \bibinfo{volume}{63}~(\bibinfo{number}{2}) (\bibinfo{year}{2022})
  \bibinfo{pages}{42}.

\bibitem[{Ding et~al.(2023)Ding, Roberts, and Schmidt}]{Ding:2022ows}
\bibinfo{author}{M.~Ding}, \bibinfo{author}{C.~D. Roberts},
  \bibinfo{author}{S.~M. Schmidt}, \bibinfo{title}{{Emergence of Hadron Mass
  and Structure}}, \bibinfo{journal}{Particles}
  \bibinfo{volume}{6}~(\bibinfo{number}{1}) (\bibinfo{year}{2023})
  \bibinfo{pages}{57--120}.

\bibitem[{Roberts(2023)}]{Roberts:2022rxm}
\bibinfo{author}{C.~D. Roberts}, \bibinfo{title}{{Origin of the Proton Mass}},
  \bibinfo{journal}{EPJ Web Conf.} \bibinfo{volume}{282} (\bibinfo{year}{2023})
  \bibinfo{pages}{01006}.

\bibitem[{Ferreira and Papavassiliou(2023)}]{Ferreira:2023fva}
\bibinfo{author}{M.~N. Ferreira}, \bibinfo{author}{J.~Papavassiliou},
  \bibinfo{title}{{Gauge Sector Dynamics in QCD}}, \bibinfo{journal}{Particles}
  \bibinfo{volume}{6}~(\bibinfo{number}{1}) (\bibinfo{year}{2023})
  \bibinfo{pages}{312--363}.

\bibitem[{Carman et~al.(2023)Carman, Gothe, Mokeev, and
  Roberts}]{Carman:2023zke}
\bibinfo{author}{D.~S. Carman}, \bibinfo{author}{R.~W. Gothe},
  \bibinfo{author}{V.~I. Mokeev}, \bibinfo{author}{C.~D. Roberts},
  \bibinfo{title}{{Nucleon Resonance Electroexcitation Amplitudes and Emergent
  Hadron Mass}}, \bibinfo{journal}{Particles}
  \bibinfo{volume}{6}~(\bibinfo{number}{1}) (\bibinfo{year}{2023})
  \bibinfo{pages}{416--439}.

\bibitem[{Raya et~al.(2024)Raya, Bashir, Binosi, Roberts, and
  Rodr\'\i{}guez-Quintero}]{Raya:2024ejx}
\bibinfo{author}{K.~Raya}, \bibinfo{author}{A.~Bashir},
  \bibinfo{author}{D.~Binosi}, \bibinfo{author}{C.~D. Roberts},
  \bibinfo{author}{J.~Rodr\'\i{}guez-Quintero}, \bibinfo{title}{{Pseudoscalar
  Mesons and Emergent Mass}}, \bibinfo{journal}{Few Body Syst.}
  \bibinfo{volume}{65}~(\bibinfo{number}{2}) (\bibinfo{year}{2024})
  \bibinfo{pages}{60}.

\bibitem[{Gell-Mann et~al.(1968)Gell-Mann, Oakes, and Renner}]{GellMann:1968rz}
\bibinfo{author}{M.~Gell-Mann}, \bibinfo{author}{R.~J. Oakes},
  \bibinfo{author}{B.~Renner}, \bibinfo{title}{{Behavior of current divergences
  under SU(3) x SU(3)}}, \bibinfo{journal}{Phys. Rev.} \bibinfo{volume}{175}
  (\bibinfo{year}{1968}) \bibinfo{pages}{2195--2199}.

\bibitem[{Gasser and Leutwyler(1984)}]{Gasser:1983yg}
\bibinfo{author}{J.~Gasser}, \bibinfo{author}{H.~Leutwyler},
  \bibinfo{title}{{Chiral Perturbation Theory to One Loop}},
  \bibinfo{journal}{Annals Phys.} \bibinfo{volume}{158} (\bibinfo{year}{1984})
  \bibinfo{pages}{142}.

\bibitem[{Llewellyn-Smith(1969)}]{LlewellynSmith:1969az}
\bibinfo{author}{C.~H. Llewellyn-Smith}, \bibinfo{title}{{A relativistic
  formulation for the quark model for mesons}}, \bibinfo{journal}{Annals Phys.}
  \bibinfo{volume}{53} (\bibinfo{year}{1969}) \bibinfo{pages}{521--558}.

\bibitem[{Dominguez(1977)}]{Dominguez:1976ut}
\bibinfo{author}{C.~A. Dominguez}, \bibinfo{title}{{Extended Partially
  Conserved Axial-Vector Current Hypothesis and Chiral Symmetry Breaking}},
  \bibinfo{journal}{Phys. Rev. D} \bibinfo{volume}{15} (\bibinfo{year}{1977})
  \bibinfo{pages}{1350--1360}.

\bibitem[{H{\"o}ll et~al.(2004)H{\"o}ll, Krassnigg, and Roberts}]{Holl:2004fr}
\bibinfo{author}{A.~H{\"o}ll}, \bibinfo{author}{A.~Krassnigg},
  \bibinfo{author}{C.~D. Roberts}, \bibinfo{title}{{Pseudoscalar meson radial
  excitations}}, \bibinfo{journal}{Phys. Rev. C} \bibinfo{volume}{70}
  (\bibinfo{year}{2004}) \bibinfo{pages}{042203(R)}.

\bibitem[{H{\"o}ll et~al.(2005)H{\"o}ll, Krassnigg, Maris, Roberts, and
  Wright}]{Holl:2005vu}
\bibinfo{author}{A.~H{\"o}ll}, \bibinfo{author}{A.~Krassnigg},
  \bibinfo{author}{P.~Maris}, \bibinfo{author}{C.~D. Roberts},
  \bibinfo{author}{S.~V. Wright}, \bibinfo{title}{{Electromagnetic properties
  of ground and excited state pseudoscalar mesons}}, \bibinfo{journal}{Phys.
  Rev. C} \bibinfo{volume}{71} (\bibinfo{year}{2005}) \bibinfo{pages}{065204}.

\bibitem[{Bhagwat et~al.(2007)Bhagwat, Krassnigg, Maris, and
  Roberts}]{Bhagwat:2006xi}
\bibinfo{author}{M.~S. Bhagwat}, \bibinfo{author}{A.~Krassnigg},
  \bibinfo{author}{P.~Maris}, \bibinfo{author}{C.~D. Roberts},
  \bibinfo{title}{{Mind the gap}}, \bibinfo{journal}{Eur. Phys. J. A}
  \bibinfo{volume}{31} (\bibinfo{year}{2007}) \bibinfo{pages}{630--637}.

\bibitem[{Ballon-Bayona et~al.(2015)Ballon-Bayona, Krein, and
  Miller}]{Ballon-Bayona:2014oma}
\bibinfo{author}{A.~Ballon-Bayona}, \bibinfo{author}{G.~Krein},
  \bibinfo{author}{C.~Miller}, \bibinfo{title}{{Decay constants of the pion and
  its excitations in holographic QCD}}, \bibinfo{journal}{Phys. Rev. D}
  \bibinfo{volume}{91} (\bibinfo{year}{2015}) \bibinfo{pages}{065024}.

\bibitem[{Jiang and Zhu(2015)}]{Jiang:2015paa}
\bibinfo{author}{J.-F. Jiang}, \bibinfo{author}{S.-L. Zhu},
  \bibinfo{title}{{Radial excitations of mesons and nucleons from QCD sum
  rules}}, \bibinfo{journal}{Phys. Rev. D}
  \bibinfo{volume}{92}~(\bibinfo{number}{7}) (\bibinfo{year}{2015})
  \bibinfo{pages}{074002}.

\bibitem[{Diehl and Hiller(2001)}]{Diehl:2001xe}
\bibinfo{author}{M.~Diehl}, \bibinfo{author}{G.~Hiller}, \bibinfo{title}{{New
  ways to explore factorization in b decays}}, \bibinfo{journal}{JHEP}
  \bibinfo{volume}{06} (\bibinfo{year}{2001}) \bibinfo{pages}{067}.

\bibitem[{McNeile and Michael(2006)}]{McNeile:2006qy}
\bibinfo{author}{C.~McNeile}, \bibinfo{author}{C.~Michael},
  \bibinfo{title}{{The decay constant of the first excited pion from lattice
  QCD}}, \bibinfo{journal}{Phys. Lett. B} \bibinfo{volume}{642}
  (\bibinfo{year}{2006}) \bibinfo{pages}{244--247}.

\bibitem[{Xu et~al.(2023)Xu, Yao, Qin, Cui, and Roberts}]{Xu:2022kng}
\bibinfo{author}{Z.-N. Xu}, \bibinfo{author}{Z.-Q. Yao}, \bibinfo{author}{S.-X.
  Qin}, \bibinfo{author}{Z.-F. Cui}, \bibinfo{author}{C.~D. Roberts},
  \bibinfo{title}{{Bethe-Salpeter kernel and properties of strange-quark
  mesons}}, \bibinfo{journal}{Eur. Phys. J. A}
  \bibinfo{volume}{59}~(\bibinfo{number}{3}) (\bibinfo{year}{2023})
  \bibinfo{pages}{39}.

\bibitem[{Li et~al.(2016)Li, Chang, Gao, Roberts, Schmidt, and
  Zong}]{Li:2016dzv}
\bibinfo{author}{B.-L. Li}, \bibinfo{author}{L.~Chang},
  \bibinfo{author}{F.~Gao}, \bibinfo{author}{C.~D. Roberts},
  \bibinfo{author}{S.~M. Schmidt}, \bibinfo{author}{H.-S. Zong},
  \bibinfo{title}{{Distribution amplitudes of radially-excited $\pi$ and $K$
  mesons}}, \bibinfo{journal}{Phys. Rev. D}
  \bibinfo{volume}{93}~(\bibinfo{number}{11}) (\bibinfo{year}{2016})
  \bibinfo{pages}{114033}.

\bibitem[{Hua et~al.(2022)}]{LatticeParton:2022zqc}
\bibinfo{author}{J.~Hua}, et~al., \bibinfo{title}{{Pion and Kaon Distribution
  Amplitudes from Lattice QCD}}, \bibinfo{journal}{Phys. Rev. Lett.}
  \bibinfo{volume}{129}~(\bibinfo{number}{13}) (\bibinfo{year}{2022})
  \bibinfo{pages}{132001}.

\bibitem[{Li(2022)}]{Li:2022qul}
\bibinfo{author}{H.-N. Li}, \bibinfo{title}{{Dispersive derivation of the pion
  distribution amplitude}}, \bibinfo{journal}{Phys. Rev. D}
  \bibinfo{volume}{106}~(\bibinfo{number}{3}) (\bibinfo{year}{2022})
  \bibinfo{pages}{034015}.

\bibitem[{Gao et~al.(2022)Gao, Hanlon, Karthik, Mukherjee, Petreczky, Scior,
  Syritsyn, and Zhao}]{Gao:2022vyh}
\bibinfo{author}{X.~Gao}, \bibinfo{author}{A.~D. Hanlon},
  \bibinfo{author}{N.~Karthik}, \bibinfo{author}{S.~Mukherjee},
  \bibinfo{author}{P.~Petreczky}, \bibinfo{author}{P.~Scior},
  \bibinfo{author}{S.~Syritsyn}, \bibinfo{author}{Y.~Zhao},
  \bibinfo{title}{{Pion distribution amplitude at the physical point using the
  leading-twist expansion of the quasi-distribution-amplitude matrix element}},
  \bibinfo{journal}{Phys. Rev. D} \bibinfo{volume}{106}~(\bibinfo{number}{7})
  (\bibinfo{year}{2022}) \bibinfo{pages}{074505}.

\bibitem[{Xing et~al.(2024{\natexlab{a}})Xing, Ding, Cui, Pimikov, Roberts, and
  Schmidt}]{Xing:2023wuk}
\bibinfo{author}{H.~Y. Xing}, \bibinfo{author}{M.~Ding}, \bibinfo{author}{Z.~F.
  Cui}, \bibinfo{author}{A.~V. Pimikov}, \bibinfo{author}{C.~D. Roberts},
  \bibinfo{author}{S.~M. Schmidt}, \bibinfo{title}{{Constraining the pion
  distribution amplitude using Drell-Yan reactions on a proton}},
  \bibinfo{journal}{Phys. Lett. B} \bibinfo{volume}{849}
  (\bibinfo{year}{2024}{\natexlab{a}}) \bibinfo{pages}{138462}.

\bibitem[{Lu et~al.(2024)Lu, Xu, Raya, Roberts, and
  Rodr\'\i{}guez-Quintero}]{Lu:2023yna}
\bibinfo{author}{Y.~Lu}, \bibinfo{author}{Y.-Z. Xu}, \bibinfo{author}{K.~Raya},
  \bibinfo{author}{C.~D. Roberts},
  \bibinfo{author}{J.~Rodr\'\i{}guez-Quintero}, \bibinfo{title}{{Pion
  distribution functions from low-order Mellin moments}},
  \bibinfo{journal}{Phys. Lett. B} \bibinfo{volume}{850} (\bibinfo{year}{2024})
  \bibinfo{pages}{138534}.

\bibitem[{Alexandrou et~al.(2024)}]{Alexandrou:2024zvn}
\bibinfo{author}{C.~Alexandrou}, et~al., \bibinfo{title}{{Quark and gluon
  momentum fractions in the pion and in the kaon -- arXiv:2405.08529
  [hep-lat]}} .

\bibitem[{Chang et~al.(2013)Chang, Cloet, Cobos-Martinez, Roberts, Schmidt, and
  Tandy}]{Chang:2013pq}
\bibinfo{author}{L.~Chang}, \bibinfo{author}{I.~C. Cloet},
  \bibinfo{author}{J.~J. Cobos-Martinez}, \bibinfo{author}{C.~D. Roberts},
  \bibinfo{author}{S.~M. Schmidt}, \bibinfo{author}{P.~C. Tandy},
  \bibinfo{title}{{Imaging dynamical chiral symmetry breaking: pion wave
  function on the light front}}, \bibinfo{journal}{Phys. Rev. Lett.}
  \bibinfo{volume}{110} (\bibinfo{year}{2013}) \bibinfo{pages}{132001}.

\bibitem[{Chang et~al.(2014)Chang, Mezrag, Moutarde, Roberts,
  Rodr{\'{\i}}guez-Quintero, and Tandy}]{Chang:2014lva}
\bibinfo{author}{L.~Chang}, \bibinfo{author}{C.~Mezrag},
  \bibinfo{author}{H.~Moutarde}, \bibinfo{author}{C.~D. Roberts},
  \bibinfo{author}{J.~Rodr{\'{\i}}guez-Quintero}, \bibinfo{author}{P.~C.
  Tandy}, \bibinfo{title}{{Basic features of the pion valence-quark
  distribution function}}, \bibinfo{journal}{Phys. Lett. B}
  \bibinfo{volume}{737} (\bibinfo{year}{2014}) \bibinfo{pages}{23--29}.

\bibitem[{Ding et~al.(2020{\natexlab{a}})Ding, Raya, Binosi, Chang, Roberts,
  and Schmidt}]{Ding:2019qlr}
\bibinfo{author}{M.~Ding}, \bibinfo{author}{K.~Raya},
  \bibinfo{author}{D.~Binosi}, \bibinfo{author}{L.~Chang},
  \bibinfo{author}{C.~D. Roberts}, \bibinfo{author}{S.~M. Schmidt},
  \bibinfo{title}{{Drawing insights from pion parton distributions}},
  \bibinfo{journal}{Chin. Phys. C (Lett.)} \bibinfo{volume}{44}
  (\bibinfo{year}{2020}{\natexlab{a}}) \bibinfo{pages}{031002}.

\bibitem[{Ding et~al.(2020{\natexlab{b}})Ding, Raya, Binosi, Chang, Roberts,
  and Schmidt}]{Ding:2019lwe}
\bibinfo{author}{M.~Ding}, \bibinfo{author}{K.~Raya},
  \bibinfo{author}{D.~Binosi}, \bibinfo{author}{L.~Chang},
  \bibinfo{author}{C.~D. Roberts}, \bibinfo{author}{S.~M. Schmidt},
  \bibinfo{title}{{Symmetry, symmetry breaking, and pion parton
  distributions}}, \bibinfo{journal}{Phys. Rev. D}
  \bibinfo{volume}{101}~(\bibinfo{number}{5})
  (\bibinfo{year}{2020}{\natexlab{b}}) \bibinfo{pages}{054014}.

\bibitem[{Diehl(2003)}]{Diehl:2003ny}
\bibinfo{author}{M.~Diehl}, \bibinfo{title}{{Generalized parton
  distributions}}, \bibinfo{journal}{Phys. Rept.} \bibinfo{volume}{388}
  (\bibinfo{year}{2003}) \bibinfo{pages}{41--277}.

\bibitem[{Nakanishi(1969)}]{Nakanishi:1969ph}
\bibinfo{author}{N.~Nakanishi}, \bibinfo{title}{{A General survey of the theory
  of the Bethe-Salpeter equation}}, \bibinfo{journal}{Prog. Theor. Phys.
  Suppl.} \bibinfo{volume}{43} (\bibinfo{year}{1969}) \bibinfo{pages}{1--81}.

\bibitem[{Ellis et~al.(1991)Ellis, Stirling, and Webber}]{Ellis:1991qj}
\bibinfo{author}{R.~K. Ellis}, \bibinfo{author}{W.~J. Stirling},
  \bibinfo{author}{B.~R. Webber}, \bibinfo{title}{{\mbox{$\;$}QCD and collider
  physics}}, \bibinfo{publisher}{Cambridge University Press, Cambridge, UK},
  \bibinfo{year}{1991}.

\bibitem[{Yin et~al.(2023)Yin, Xu, Cui, Roberts, and
  Rodr\'\i{}guez-Quintero}]{Yin:2023dbw}
\bibinfo{author}{P.-L. Yin}, \bibinfo{author}{Y.-Z. Xu}, \bibinfo{author}{Z.-F.
  Cui}, \bibinfo{author}{C.~D. Roberts},
  \bibinfo{author}{J.~Rodr\'\i{}guez-Quintero}, \bibinfo{title}{{All-Orders
  Evolution of Parton Distributions: Principle, Practice, and Predictions}},
  \bibinfo{journal}{Chin. Phys. Lett. \emph{Express}}
  \bibinfo{volume}{40}~(\bibinfo{number}{9}) (\bibinfo{year}{2023})
  \bibinfo{pages}{091201}.

\bibitem[{Grunberg(1980)}]{Grunberg:1980ja}
\bibinfo{author}{G.~Grunberg}, \bibinfo{title}{{Renormalization Group Improved
  Perturbative QCD}}, \bibinfo{journal}{Phys. Lett. B} \bibinfo{volume}{95}
  (\bibinfo{year}{1980}) \bibinfo{pages}{70}, \bibinfo{note}{[Erratum: Phys.
  Lett. B 110, 501 (1982)]}.

\bibitem[{Grunberg(1984)}]{Grunberg:1982fw}
\bibinfo{author}{G.~Grunberg}, \bibinfo{title}{{Renormalization Scheme
  Independent QCD and QED: The Method of Effective Charges}},
  \bibinfo{journal}{Phys. Rev. D} \bibinfo{volume}{29} (\bibinfo{year}{1984})
  \bibinfo{pages}{2315}.

\bibitem[{Deur et~al.(2024)Deur, Brodsky, and Roberts}]{Deur:2023dzc}
\bibinfo{author}{A.~Deur}, \bibinfo{author}{S.~J. Brodsky},
  \bibinfo{author}{C.~D. Roberts}, \bibinfo{title}{{QCD Running Couplings and
  Effective Charges}}, \bibinfo{journal}{Prog. Part. Nucl. Phys.}
  \bibinfo{volume}{134} (\bibinfo{year}{2024}) \bibinfo{pages}{104081}.

\bibitem[{Dokshitzer(1977)}]{Dokshitzer:1977sg}
\bibinfo{author}{Y.~L. Dokshitzer}, \bibinfo{title}{Calculation of the
  Structure Functions for Deep Inelastic Scattering and $e^+$ $e^-$
  Annihilation by Perturbation Theory in Quantum Chromodynamics. ({\mbox {I}n
  {R}ussian})}, \bibinfo{journal}{Sov. Phys. JETP} \bibinfo{volume}{46}
  (\bibinfo{year}{1977}) \bibinfo{pages}{641--653}.

\bibitem[{Gribov and Lipatov(1971)}]{Gribov:1971zn}
\bibinfo{author}{V.~N. Gribov}, \bibinfo{author}{L.~N. Lipatov},
  \bibinfo{title}{{Deep inelastic electron scattering in perturbation theory}},
  \bibinfo{journal}{Phys. Lett. B} \bibinfo{volume}{37} (\bibinfo{year}{1971})
  \bibinfo{pages}{78--80}.

\bibitem[{Lipatov(1975)}]{Lipatov:1974qm}
\bibinfo{author}{L.~N. Lipatov}, \bibinfo{title}{{The parton model and
  perturbation theory}}, \bibinfo{journal}{Sov. J. Nucl. Phys.}
  \bibinfo{volume}{20} (\bibinfo{year}{1975}) \bibinfo{pages}{94--102}.

\bibitem[{Altarelli and Parisi(1977)}]{Altarelli:1977zs}
\bibinfo{author}{G.~Altarelli}, \bibinfo{author}{G.~Parisi},
  \bibinfo{title}{{Asymptotic Freedom in Parton Language}},
  \bibinfo{journal}{Nucl. Phys. B} \bibinfo{volume}{126} (\bibinfo{year}{1977})
  \bibinfo{pages}{298--318}.

\bibitem[{Binosi et~al.(2015)Binosi, Chang, Papavassiliou, and
  Roberts}]{Binosi:2014aea}
\bibinfo{author}{D.~Binosi}, \bibinfo{author}{L.~Chang},
  \bibinfo{author}{J.~Papavassiliou}, \bibinfo{author}{C.~D. Roberts},
  \bibinfo{title}{{Bridging a gap between continuum-QCD and \emph{ab initio}
  predictions of hadron observables}}, \bibinfo{journal}{Phys. Lett. B}
  \bibinfo{volume}{742} (\bibinfo{year}{2015}) \bibinfo{pages}{183--188}.

\bibitem[{Binosi et~al.(2017)Binosi, Mezrag, Papavassiliou, Roberts, and
  Rodr{\'i}guez-Quintero}]{Binosi:2016nme}
\bibinfo{author}{D.~Binosi}, \bibinfo{author}{C.~Mezrag},
  \bibinfo{author}{J.~Papavassiliou}, \bibinfo{author}{C.~D. Roberts},
  \bibinfo{author}{J.~Rodr{\'i}guez-Quintero},
  \bibinfo{title}{{Process-independent strong running coupling}},
  \bibinfo{journal}{Phys. Rev. D} \bibinfo{volume}{96} (\bibinfo{year}{2017})
  \bibinfo{pages}{054026}.

\bibitem[{Cui et~al.(2020{\natexlab{a}})Cui, Zhang, Binosi, de~Soto, Mezrag,
  Papavassiliou, Roberts, Rodr{\'{\i}}guez-Quintero, Segovia, and
  Zafeiropoulos}]{Cui:2019dwv}
\bibinfo{author}{Z.-F. Cui}, \bibinfo{author}{J.-L. Zhang},
  \bibinfo{author}{D.~Binosi}, \bibinfo{author}{F.~de~Soto},
  \bibinfo{author}{C.~Mezrag}, \bibinfo{author}{J.~Papavassiliou},
  \bibinfo{author}{C.~D. Roberts},
  \bibinfo{author}{J.~Rodr{\'{\i}}guez-Quintero}, \bibinfo{author}{J.~Segovia},
  \bibinfo{author}{S.~Zafeiropoulos}, \bibinfo{title}{{Effective charge from
  lattice QCD}}, \bibinfo{journal}{Chin. Phys. C} \bibinfo{volume}{44}
  (\bibinfo{year}{2020}{\natexlab{a}}) \bibinfo{pages}{083102}.

\bibitem[{Xing et~al.(2024{\natexlab{b}})Xing, Yao, Li, Binosi, Cui, and
  Roberts}]{Xing:2023pms}
\bibinfo{author}{H.~Y. Xing}, \bibinfo{author}{Z.~Q. Yao},
  \bibinfo{author}{B.~L. Li}, \bibinfo{author}{D.~Binosi},
  \bibinfo{author}{Z.~F. Cui}, \bibinfo{author}{C.~D. Roberts},
  \bibinfo{title}{{Developing predictions for pion fragmentation functions}},
  \bibinfo{journal}{Eur. Phys. J. C} \bibinfo{volume}{84}~(\bibinfo{number}{1})
  (\bibinfo{year}{2024}{\natexlab{b}}) \bibinfo{pages}{82}.

\bibitem[{Cui et~al.(2020{\natexlab{b}})Cui, Ding, Gao, Raya, Binosi, Chang,
  Roberts, Rodr\'{\i}guez-Quintero, and Schmidt}]{Cui:2020tdf}
\bibinfo{author}{Z.-F. Cui}, \bibinfo{author}{M.~Ding},
  \bibinfo{author}{F.~Gao}, \bibinfo{author}{K.~Raya},
  \bibinfo{author}{D.~Binosi}, \bibinfo{author}{L.~Chang},
  \bibinfo{author}{C.~D. Roberts},
  \bibinfo{author}{J.~Rodr\'{\i}guez-Quintero}, \bibinfo{author}{S.~M.
  Schmidt}, \bibinfo{title}{{Kaon and pion parton distributions}},
  \bibinfo{journal}{Eur. Phys. J. C} \bibinfo{volume}{80}
  (\bibinfo{year}{2020}{\natexlab{b}}) \bibinfo{pages}{1064}.

\bibitem[{Chang et~al.(2022)Chang, Gao, and Roberts}]{Chang:2022jri}
\bibinfo{author}{L.~Chang}, \bibinfo{author}{F.~Gao}, \bibinfo{author}{C.~D.
  Roberts}, \bibinfo{title}{{Parton distributions of light quarks and
  antiquarks in the proton}}, \bibinfo{journal}{Phys. Lett. B}
  \bibinfo{volume}{829} (\bibinfo{year}{2022}) \bibinfo{pages}{137078}.

\bibitem[{Lu et~al.(2022)Lu, Chang, Raya, Roberts, and
  Rodr\'\i{}guez-Quintero}]{Lu:2022cjx}
\bibinfo{author}{Y.~Lu}, \bibinfo{author}{L.~Chang}, \bibinfo{author}{K.~Raya},
  \bibinfo{author}{C.~D. Roberts},
  \bibinfo{author}{J.~Rodr\'\i{}guez-Quintero}, \bibinfo{title}{{Proton and
  pion distribution functions in counterpoint}}, \bibinfo{journal}{Phys. Lett.
  B} \bibinfo{volume}{830} (\bibinfo{year}{2022}) \bibinfo{pages}{137130}.

\bibitem[{Cheng et~al.(2023)Cheng, Yu, Xing, Chen, Cui, and
  Roberts}]{Cheng:2023kmt}
\bibinfo{author}{P.~Cheng}, \bibinfo{author}{Y.~Yu}, \bibinfo{author}{H.-Y.
  Xing}, \bibinfo{author}{C.~Chen}, \bibinfo{author}{Z.-F. Cui},
  \bibinfo{author}{C.~D. Roberts}, \bibinfo{title}{{Perspective on polarised
  parton distribution functions and proton spin}}, \bibinfo{journal}{Phys.
  Lett. B} \bibinfo{volume}{844} (\bibinfo{year}{2023})
  \bibinfo{pages}{138074}.

\bibitem[{Yu et~al.(2024)Yu, Cheng, Xing, Gao, and Roberts}]{Yu:2024qsd}
\bibinfo{author}{Y.~Yu}, \bibinfo{author}{P.~Cheng}, \bibinfo{author}{H.-Y.
  Xing}, \bibinfo{author}{F.~Gao}, \bibinfo{author}{C.~D. Roberts},
  \bibinfo{title}{{Contact interaction study of proton parton distributions}},
  \bibinfo{journal}{Eur. Phys. J. C} \bibinfo{volume}{84}~(\bibinfo{number}{7})
  (\bibinfo{year}{2024}) \bibinfo{pages}{739}.

\bibitem[{Yao et~al.(2024)Yao, Xu, Binosi, Cui, Ding, Raya, Roberts,
  Rodr\'\i{}guez-Quintero, and Schmidt}]{Yao:2024ixu}
\bibinfo{author}{Z.~Q. Yao}, \bibinfo{author}{Y.~Z. Xu},
  \bibinfo{author}{D.~Binosi}, \bibinfo{author}{Z.~F. Cui},
  \bibinfo{author}{M.~Ding}, \bibinfo{author}{K.~Raya}, \bibinfo{author}{C.~D.
  Roberts}, \bibinfo{author}{J.~Rodr\'\i{}guez-Quintero},
  \bibinfo{author}{S.~M. Schmidt}, \bibinfo{title}{{Nucleon Gravitational Form
  Factors -- arXiv:2409.15547 [hep-ph]}} .

\bibitem[{Yu and Roberts(2024)}]{Yu:2024ovn}
\bibinfo{author}{Y.~Yu}, \bibinfo{author}{C.~D. Roberts},
  \bibinfo{title}{{Impressions of Parton Distribution Functions}},
  \bibinfo{journal}{Chin. Phys. Lett.} \bibinfo{volume}{41}
  (\bibinfo{year}{2024}) \bibinfo{pages}{121202}.

\bibitem[{Roberts and Williams(1994)}]{Roberts:1994dr}
\bibinfo{author}{C.~D. Roberts}, \bibinfo{author}{A.~G. Williams},
  \bibinfo{title}{{Dyson-Schwinger equations and their application to hadronic
  physics}}, \bibinfo{journal}{Prog. Part. Nucl. Phys.} \bibinfo{volume}{33}
  (\bibinfo{year}{1994}) \bibinfo{pages}{477--575}.

\bibitem[{Munczek(1995)}]{Munczek:1994zz}
\bibinfo{author}{H.~J. Munczek}, \bibinfo{title}{{Dynamical chiral symmetry
  breaking, Goldstone's theorem and the consistency of the Schwinger-Dyson and
  Bethe-Salpeter Equations}}, \bibinfo{journal}{Phys. Rev. D}
  \bibinfo{volume}{52} (\bibinfo{year}{1995}) \bibinfo{pages}{4736--4740}.

\bibitem[{Bender et~al.(1996)Bender, Roberts, and von Smekal}]{Bender:1996bb}
\bibinfo{author}{A.~Bender}, \bibinfo{author}{C.~D. Roberts},
  \bibinfo{author}{L.~von Smekal}, \bibinfo{title}{{Goldstone Theorem and
  Diquark Confinement Beyond Rainbow- Ladder Approximation}},
  \bibinfo{journal}{Phys. Lett. B} \bibinfo{volume}{380} (\bibinfo{year}{1996})
  \bibinfo{pages}{7--12}.

\bibitem[{Qin and Roberts(2021)}]{Qin:2020jig}
\bibinfo{author}{S.-X. Qin}, \bibinfo{author}{C.~D. Roberts},
  \bibinfo{title}{{Resolving the Bethe-Salpeter kernel}},
  \bibinfo{journal}{Chin. Phys. Lett. \emph{Express}}
  \bibinfo{volume}{38}~(\bibinfo{number}{7}) (\bibinfo{year}{2021})
  \bibinfo{pages}{071201}.

\bibitem[{Navas et~al.(2024)}]{ParticleDataGroup:2024cfk}
\bibinfo{author}{S.~Navas}, et~al., \bibinfo{title}{{Review of particle
  physics}}, \bibinfo{journal}{Phys. Rev. D}
  \bibinfo{volume}{110}~(\bibinfo{number}{3}) (\bibinfo{year}{2024})
  \bibinfo{pages}{030001}.

\bibitem[{Gell-Mann and Low(1954)}]{GellMann:1954fq}
\bibinfo{author}{M.~Gell-Mann}, \bibinfo{author}{F.~E. Low},
  \bibinfo{title}{{Quantum electrodynamics at small distances}},
  \bibinfo{journal}{Phys. Rev.} \bibinfo{volume}{95} (\bibinfo{year}{1954})
  \bibinfo{pages}{1300--1312}.

\bibitem[{Brodsky et~al.(2024)Brodsky, Deur, and Roberts}]{Brodsky:2024zev}
\bibinfo{author}{S.~J. Brodsky}, \bibinfo{author}{A.~Deur},
  \bibinfo{author}{C.~D. Roberts}, \bibinfo{title}{{The Secret to the Strongest
  Force in the Universe}}, \bibinfo{journal}{Sci. Am.} \bibinfo{volume}{5
  (May)} (\bibinfo{year}{2024}) \bibinfo{pages}{32--39}.

\bibitem[{Chang et~al.(2009)Chang, Liu, Roberts, Shi, Sun, and
  Zong}]{Chang:2008ec}
\bibinfo{author}{L.~Chang}, \bibinfo{author}{Y.-X. Liu}, \bibinfo{author}{C.~D.
  Roberts}, \bibinfo{author}{Y.-M. Shi}, \bibinfo{author}{W.-M. Sun},
  \bibinfo{author}{H.-S. Zong}, \bibinfo{title}{{Chiral susceptibility and the
  scalar Ward identity}}, \bibinfo{journal}{Phys. Rev. C} \bibinfo{volume}{79}
  (\bibinfo{year}{2009}) \bibinfo{pages}{035209}.

\bibitem[{Qin et~al.(2011)Qin, Chang, Liu, Roberts, and Wilson}]{Qin:2011dd}
\bibinfo{author}{S.-X. Qin}, \bibinfo{author}{L.~Chang}, \bibinfo{author}{Y.-X.
  Liu}, \bibinfo{author}{C.~D. Roberts}, \bibinfo{author}{D.~J. Wilson},
  \bibinfo{title}{{Interaction model for the gap equation}},
  \bibinfo{journal}{Phys. Rev. C} \bibinfo{volume}{84} (\bibinfo{year}{2011})
  \bibinfo{pages}{042202(R)}.

\bibitem[{Fischer and Williams(2009)}]{Fischer:2009jm}
\bibinfo{author}{C.~S. Fischer}, \bibinfo{author}{R.~Williams},
  \bibinfo{title}{{Probing the gluon self-interaction in light mesons}},
  \bibinfo{journal}{Phys. Rev. Lett.} \bibinfo{volume}{103}
  (\bibinfo{year}{2009}) \bibinfo{pages}{122001}.

\bibitem[{Chang and Roberts(2009)}]{Chang:2009zb}
\bibinfo{author}{L.~Chang}, \bibinfo{author}{C.~D. Roberts},
  \bibinfo{title}{{Sketching the Bethe-Salpeter kernel}},
  \bibinfo{journal}{Phys. Rev. Lett.} \bibinfo{volume}{103}
  (\bibinfo{year}{2009}) \bibinfo{pages}{081601}.

\bibitem[{Qin et~al.(2012)Qin, Chang, Liu, Roberts, and Wilson}]{Qin:2011xq}
\bibinfo{author}{S.-X. Qin}, \bibinfo{author}{L.~Chang}, \bibinfo{author}{Y.-X.
  Liu}, \bibinfo{author}{C.~D. Roberts}, \bibinfo{author}{D.~J. Wilson},
  \bibinfo{title}{{Investigation of rainbow-ladder truncation for excited and
  exotic mesons}}, \bibinfo{journal}{Phys. Rev. C} \bibinfo{volume}{85}
  (\bibinfo{year}{2012}) \bibinfo{pages}{035202}.

\bibitem[{Chang and Roberts(2012)}]{Chang:2011ei}
\bibinfo{author}{L.~Chang}, \bibinfo{author}{C.~D. Roberts},
  \bibinfo{title}{{Tracing masses of ground-state light-quark mesons}},
  \bibinfo{journal}{Phys. Rev. C} \bibinfo{volume}{85} (\bibinfo{year}{2012})
  \bibinfo{pages}{052201(R)}.

\bibitem[{Chang et~al.(2011)Chang, Liu, and Roberts}]{Chang:2010hb}
\bibinfo{author}{L.~Chang}, \bibinfo{author}{Y.-X. Liu}, \bibinfo{author}{C.~D.
  Roberts}, \bibinfo{title}{{Dressed-quark anomalous magnetic moments}},
  \bibinfo{journal}{Phys. Rev. Lett.} \bibinfo{volume}{106}
  (\bibinfo{year}{2011}) \bibinfo{pages}{072001}.

\bibitem[{Maris and Roberts(1997)}]{Maris:1997tm}
\bibinfo{author}{P.~Maris}, \bibinfo{author}{C.~D. Roberts},
  \bibinfo{title}{{{$\pi$} and {$K$} meson Bethe-Salpeter amplitudes}},
  \bibinfo{journal}{Phys. Rev. C} \bibinfo{volume}{56} (\bibinfo{year}{1997})
  \bibinfo{pages}{3369--3383}.

\bibitem[{Maris and Tandy(2006)}]{Maris:2005tt}
\bibinfo{author}{P.~Maris}, \bibinfo{author}{P.~C. Tandy}, \bibinfo{title}{{QCD
  modeling of hadron physics}}, \bibinfo{journal}{Nucl. Phys. Proc. Suppl.}
  \bibinfo{volume}{161} (\bibinfo{year}{2006}) \bibinfo{pages}{136--152}.

\bibitem[{Krassnigg(2008)}]{Krassnigg:2009gd}
\bibinfo{author}{A.~Krassnigg}, \bibinfo{title}{{Excited mesons in a
  Bethe-Salpeter approach}}, \bibinfo{journal}{PoS}
  \bibinfo{volume}{CONFINEMENT\,8} (\bibinfo{year}{2008}) \bibinfo{pages}{075}.

\bibitem[{Schlessinger and Schwartz(1966)}]{Schlessinger:1966zz}
\bibinfo{author}{L.~Schlessinger}, \bibinfo{author}{C.~Schwartz},
  \bibinfo{title}{{Analyticity as a Useful Computation Tool}},
  \bibinfo{journal}{Phys. Rev. Lett.} \bibinfo{volume}{16}
  (\bibinfo{year}{1966}) \bibinfo{pages}{1173--1174}.

\bibitem[{Schlessinger(1968)}]{PhysRev.167.1411}
\bibinfo{author}{L.~Schlessinger}, \bibinfo{title}{Use of Analyticity in the
  Calculation of Nonrelativistic Scattering Amplitudes},
  \bibinfo{journal}{Phys. Rev.} \bibinfo{volume}{167} (\bibinfo{year}{1968})
  \bibinfo{pages}{1411--1423}.

\bibitem[{Tripolt et~al.(2017)Tripolt, Haritan, Wambach, and
  Moiseyev}]{Tripolt:2016cya}
\bibinfo{author}{R.~A. Tripolt}, \bibinfo{author}{I.~Haritan},
  \bibinfo{author}{J.~Wambach}, \bibinfo{author}{N.~Moiseyev},
  \bibinfo{title}{{Threshold energies and poles for hadron physical problems by
  a model-independent universal algorithm}}, \bibinfo{journal}{Phys. Lett. B}
  \bibinfo{volume}{774} (\bibinfo{year}{2017}) \bibinfo{pages}{411--416}.

\bibitem[{Cui et~al.(2022{\natexlab{a}})Cui, Binosi, Roberts, and
  Schmidt}]{Cui:2022fyr}
\bibinfo{author}{Z.-F. Cui}, \bibinfo{author}{D.~Binosi},
  \bibinfo{author}{C.~D. Roberts}, \bibinfo{author}{S.~M. Schmidt},
  \bibinfo{title}{{Hadron and light nucleus radii from electron scattering}},
  \bibinfo{journal}{Chin. Phys. C} \bibinfo{volume}{46}~(\bibinfo{number}{12})
  (\bibinfo{year}{2022}{\natexlab{a}}) \bibinfo{pages}{122001}.

\bibitem[{Cui et~al.(2022{\natexlab{b}})Cui, Ding, Morgado, Raya, Binosi,
  Chang, Papavassiliou, Roberts, Rodr\'\i{}guez-Quintero, and
  Schmidt}]{Cui:2021mom}
\bibinfo{author}{Z.~F. Cui}, \bibinfo{author}{M.~Ding}, \bibinfo{author}{J.~M.
  Morgado}, \bibinfo{author}{K.~Raya}, \bibinfo{author}{D.~Binosi},
  \bibinfo{author}{L.~Chang}, \bibinfo{author}{J.~Papavassiliou},
  \bibinfo{author}{C.~D. Roberts},
  \bibinfo{author}{J.~Rodr\'\i{}guez-Quintero}, \bibinfo{author}{S.~M.
  Schmidt}, \bibinfo{title}{{Concerning pion parton distributions}},
  \bibinfo{journal}{Eur. Phys. J. A} \bibinfo{volume}{58}~(\bibinfo{number}{1})
  (\bibinfo{year}{2022}{\natexlab{b}}) \bibinfo{pages}{10}.

\bibitem[{Cui et~al.(2022{\natexlab{c}})Cui, Ding, Morgado, Raya, Binosi,
  Chang, De~Soto, Roberts, Rodr\'\i{}guez-Quintero, and Schmidt}]{Cui:2022bxn}
\bibinfo{author}{Z.~F. Cui}, \bibinfo{author}{M.~Ding}, \bibinfo{author}{J.~M.
  Morgado}, \bibinfo{author}{K.~Raya}, \bibinfo{author}{D.~Binosi},
  \bibinfo{author}{L.~Chang}, \bibinfo{author}{F.~De~Soto},
  \bibinfo{author}{C.~D. Roberts},
  \bibinfo{author}{J.~Rodr\'\i{}guez-Quintero}, \bibinfo{author}{S.~M.
  Schmidt}, \bibinfo{title}{{Emergence of pion parton distributions}},
  \bibinfo{journal}{Phys. Rev. D} \bibinfo{volume}{105}~(\bibinfo{number}{9})
  (\bibinfo{year}{2022}{\natexlab{c}}) \bibinfo{pages}{L091502}.

\bibitem[{Raya et~al.(2022)Raya, Cui, Chang, Morgado, Roberts, and
  Rodr{\'{\i}}guez-Quintero}]{Raya:2021zrz}
\bibinfo{author}{K.~Raya}, \bibinfo{author}{Z.-F. Cui},
  \bibinfo{author}{L.~Chang}, \bibinfo{author}{J.-M. Morgado},
  \bibinfo{author}{C.~D. Roberts},
  \bibinfo{author}{J.~Rodr{\'{\i}}guez-Quintero}, \bibinfo{title}{{Revealing
  pion and kaon structure via generalised parton distributions}},
  \bibinfo{journal}{Chin. Phys. C} \bibinfo{volume}{46}~(\bibinfo{number}{26})
  (\bibinfo{year}{2022}) \bibinfo{pages}{013105}.

\bibitem[{Gao et~al.(2021)Gao, Karthik, Mukherjee, Petreczky, Syritsyn, and
  Zhao}]{Gao:2021hvs}
\bibinfo{author}{X.~Gao}, \bibinfo{author}{N.~Karthik},
  \bibinfo{author}{S.~Mukherjee}, \bibinfo{author}{P.~Petreczky},
  \bibinfo{author}{S.~Syritsyn}, \bibinfo{author}{Y.~Zhao},
  \bibinfo{title}{{Towards studying the structural differences between the pion
  and its radial excitation}}, \bibinfo{journal}{Phys. Rev. D}
  \bibinfo{volume}{103}~(\bibinfo{number}{9}) (\bibinfo{year}{2021})
  \bibinfo{pages}{094510}.

\end{thebibliography}

\end{document}